\documentclass[twocolumn,preprintnumbers,amsfonts,amsmath,amssymb,aps,prb,floatfix,showpacs,longbibliography]{revtex4-2}
\usepackage{graphicx} % needed for figures
\usepackage[leftmargin=2em,rightmargin=2em,vskip=2pt]{quoting}

\usepackage[dvipsnames]{xcolor}
\definecolor{darkgreen}{hsb}{.333,1,.5}
\definecolor{darkblue}{hsb}{.667,1,.75}
\definecolor{darkbrown}{hsb}{0,.5,.5}
\usepackage{hyperref}	% add hypertext capabilities
\hypersetup{
    colorlinks=true,
    linkcolor=darkbrown,	% equations
    urlcolor=darkblue,		% urls
    citecolor=darkgreen,	% references
    breaklinks=true,
}
\DeclareMathOperator{\acos}{acos}
\DeclareMathOperator{\acot}{acot}
\DeclareMathOperator{\atan}{atan}
\DeclareMathOperator{\sign}{sign}

\begin{document}

\title{Black hole singularity is a surface not a point}

%\author{Author}
%\email{email}
%\homepage{http://jila.colorado.edu/~ajsh/}
%\affiliation{Affiliation}

\author{Andrew J. S. Hamilton}
\email{Andrew.Hamilton@colorado.edu}
\homepage{http://jila.colorado.edu/~ajsh/}
\affiliation{
JILA and Dept.\ Astrophysical \& Planetary Sciences,
Box 440, U.\ Colorado Boulder, CO 80309, USA
}

\author{Tyler McMaken}
\email{tcmcmaken@umary.edu}
%\homepage{https://www.umary.edu/about/directory/tyler-mcmaken-phd}
\affiliation{
Dept.\ of Mathematics and Physics,
University of Mary, 7500 University Drive, Bismarck, ND 58504, USA
}

\newcommand{\dd}{d}
\newcommand{\ee}{e}

\newcommand{\unit}[1]{\, {\rm #1}}
\newcommand{\Msun}{{\textrm{M}_\odot}}
\newcommand{\Mbh}{{\textrm{M}_\bullet}}

\newcommand{\diag}{\textrm{diag}}
\newcommand{\obs}{\textrm{obs}}
\newcommand{\pn}{\textrm{pn}}

\newcommand{\bJ}{\bm{J}}
\newcommand{\bL}{\bm{L}}

\newcommand{\KCarter}{{\cal K}}

%--------------------
% FIG
\newcommand{\schwpenrosefig}{
    \begin{figure}[t!]
    \begin{center}
    \includegraphics[scale=.83]{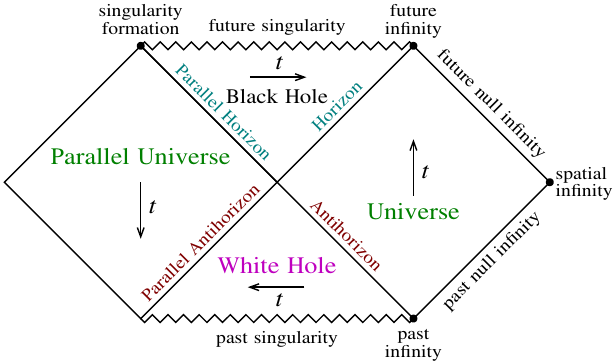}
    \caption[Penrose spacetime diagram]{
    \label{schwpenrose}
Penrose diagram of the analytically extended Schwarzschild geometry.
A Penrose diagram \cite{Penrose:1964ge} is a spacetime diagram
constructed from radial and time coordinates arranged
such that radially moving light,
whether outgoing ($\nearrow$) or ingoing ($\nwarrow$),
always moves at 45$^\circ$,
and spatial and temporal infinity are mapped to a finite position.
%A Penrose diagram clarifies the causal structure of the geometry,
%since timelike worldlines must evolve generally upward,
%at less than $45^\circ$ from vertical.
The analytically extended geometry contains
not only a Universe and a Black Hole,
but also a Parallel Universe and a White Hole.
The Parallel Universe and White Hole are time-reversed versions
of the Universe and Black Hole.
In a real black hole formed from gravitational collapse,
there is no Parallel Universe or White Hole,
and the Antihorizon and Parallel Horizon
are replaced by the dimming, redshifting surface of whatever
collapsed into the black hole long ago.
The arrows show the direction of the Schwarzschild time coordinate $t$
in each of the regions.
Inside the Black Hole and White Hole, the time coordinate $t$ is spacelike.
The endpoints of the future singularity are labeled
from the perspective of observers in the Universe,
``singularity formation'' and ``future infinity.''
    }
    \end{center}
    \end{figure}
}

%%--------------------
%% FIG
%\newcommand{\penroseknfig}{
%    \begin{figure}[tbp!]
%    \begin{center}
%    \leavevmode
%    \includegraphics[scale=1]{penrose_kn}
%    \caption[Penrose diagram of the Kerr-Newman geometry]{
%    \label{penrosekn}
%Penrose diagram of the Kerr-Newman geometry.
%The diagram is similar to that of the Reissner-Nordstr\"om geometry,
%except that it is possible to pass through the disk at $r = 0$
%from the Wormhole region into the Antiverse region.
%This Penrose diagram, which represents a slice
%at fixed $\theta$ and $\phi$,
%does not capture the full richness of the geometry,
%which contains closed timelike curves in a torus
%around the ring singularity,
%the sisytube.
%    }
%    \end{center}
%    \end{figure}
%}

%--------------------
% FIG
\newcommand{\penrosemassinflationfig}{
    \begin{figure}[t!]
    \begin{center}
    \leavevmode
    \includegraphics[width=3.4in]{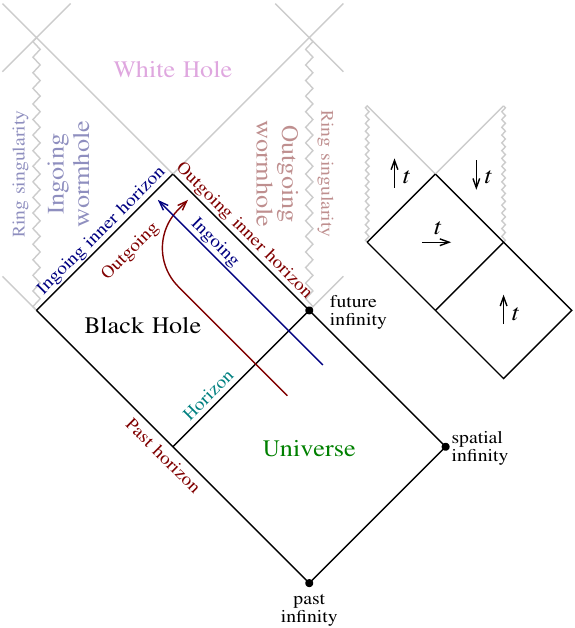}
    \caption{
Penrose diagram of the analytically extended Kerr geometry.
After passing through the Horizon into the Black Hole region where the global time coordinate $t$ becomes spacelike, geodesics may continue through either the ingoing or outgoing portions of the inner horizon.
Beyond each inner horizon is a Wormhole where coordinate time $t$ reverts
to being timelike, and an infinitely gravitationally repulsive ring singularity
is visible.
The Wormholes exit to a White Hole and more.
These parts of the analytic continuation are shown in ghostly lines and labels;
they are prevented from coming into existence
by the Poisson-Israel \cite{Poisson:1990eh}
mass inflation instability described in \S\ref{massinflation-sec}.
%Penrose diagram illustrating why Kerr black holes
%are subject to the Poisson-Israel \cite{Poisson:1990eh}
%mass inflation instability.
%Outgoing (prograde)
%and
%ingoing (retrograde)
%streams just outside the inner horizon must pass through
%separate outgoing and ingoing inner horizons
%into causally separated pieces of spacetime where the
%timelike time coordinate $t$ goes in opposite directions.
%To accomplish this, the outgoing and ingoing streams must
%exceed the speed of light through each other,
%which physically they cannot do.
%In a real black hole,
%the energy-momentum of
%hyper-relativistically counter-streaming outgoing and ingoing streams
%just above the inner horizon back-reacts on the geometry,
%leading to inflation and collapse,
%and cutting off the analytic continuations
%through the inner horizons of the Kerr geometry.
%Parts of the analytic continuation
%(wormholes, ring singularities, white hole)
%are shown in ghostly lines and labels;
%these parts never come into existence.
%The inflationary instability is driven by the pressure of the
%relativistic counter-streaming between ingoing and outgoing streams.
The inset shows the direction of coordinate time $t$
in the various regions.
Proper time of course always increases upward
in a Penrose diagram.
    }
    \label{penrosemassinflation}
    \end{center}
    \end{figure}
}

%--------------------
% FIG
\newcommand{\schwJoovisfig}{
    \begin{figure*}[t!]
    \begin{center}
    \leavevmode
    \includegraphics[scale=1]{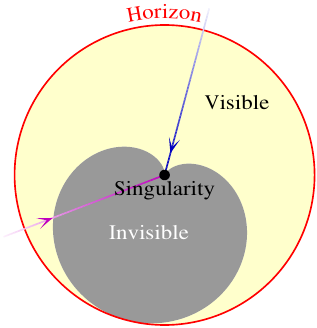}
    \hspace{4em}
    \includegraphics[scale=1]{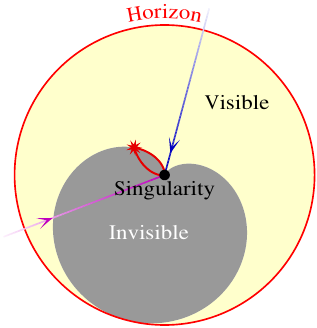}
    \caption[The singularity is not a point]{
    \label{schwJoovis}
Left: the light yellow shaded region shows the region
visible to a blue infaller who falls radially to the
singularity of a Schwarzschild black hole;
the dark gray shaded region shows the region that remains invisible
to the infaller.
The boundary of the invisible region has the shape of a cardioid,
equation~(\ref{rschwJoo}).
If another, purple, infaller falls along a different radial direction,
the two infallers not only fail to meet at the singularity,
they lose causal contact with each other
already some distance from the singularity.
Since the two infallers fall to two causally disconnected places,
the singularity cannot be a point.
Right: the same, showing in red the shortest causal path joining the two
infallers asymptotically near the singularity.
The shortest causal path is a pair of light rays that start at the
starred point, move in opposite angular directions,
and reach the infallers asymptotically near the singularity.
The shortest causal path remains nonzero even though the spatial distance
between the infallers tends to zero.
    }
    \end{center}
    \end{figure*}
}

%--------------------
% FIG
\newcommand{\schwvizfig}{
    \begin{figure*}[tp!]
    \begin{center}
    \leavevmode
    \includegraphics[scale=.125]{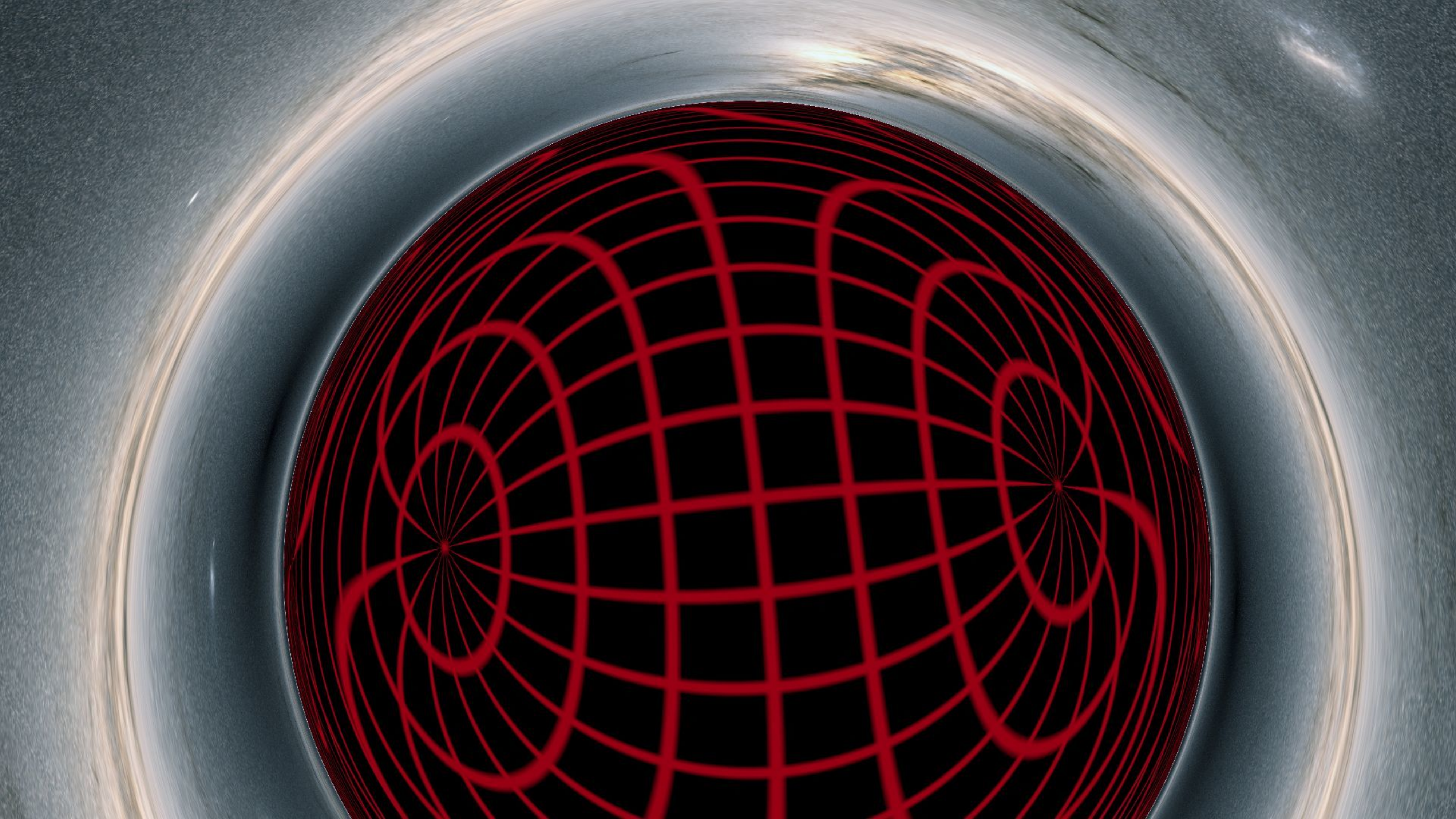}
    \includegraphics[scale=.125]{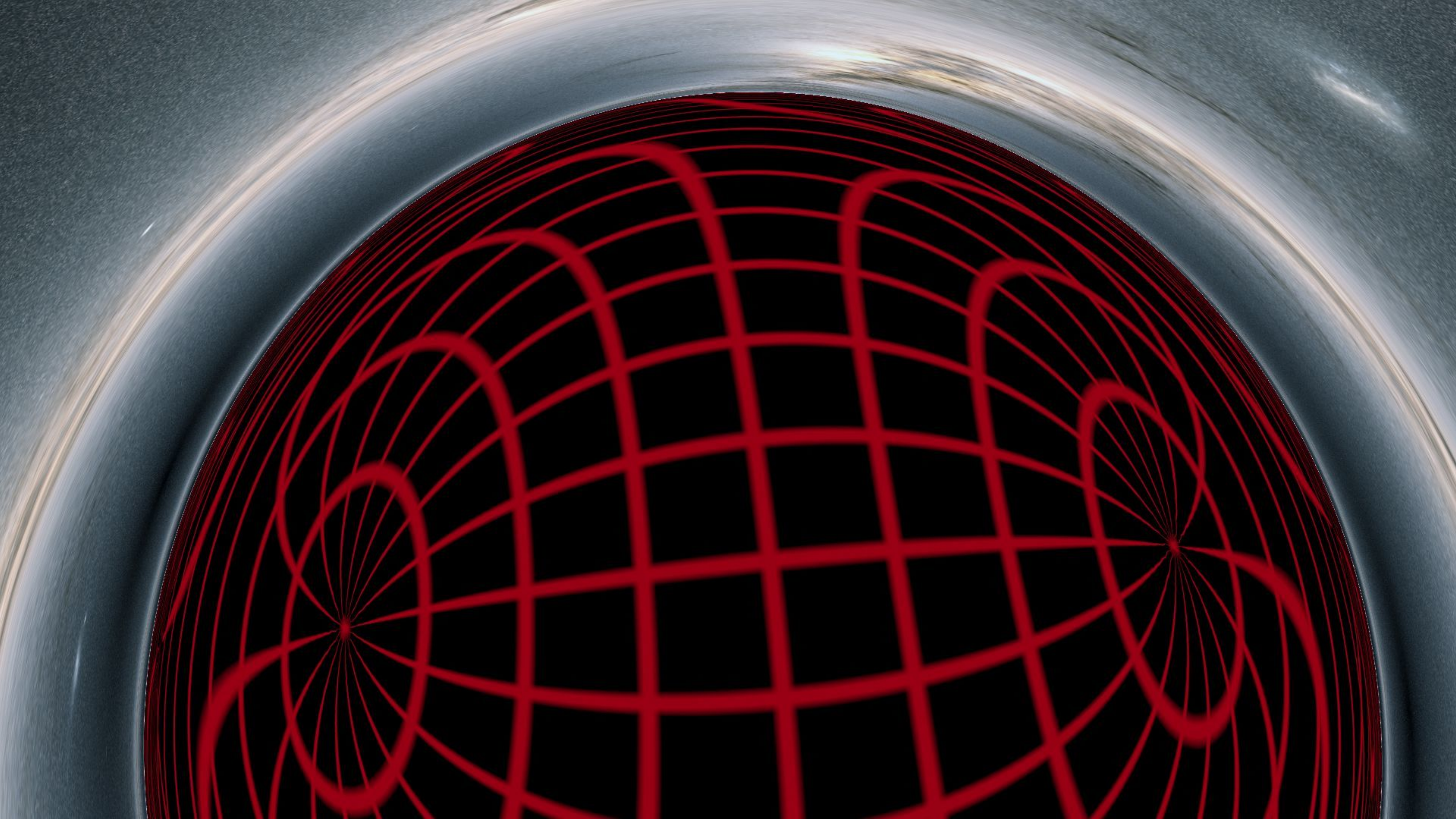}
    \includegraphics[scale=.125]{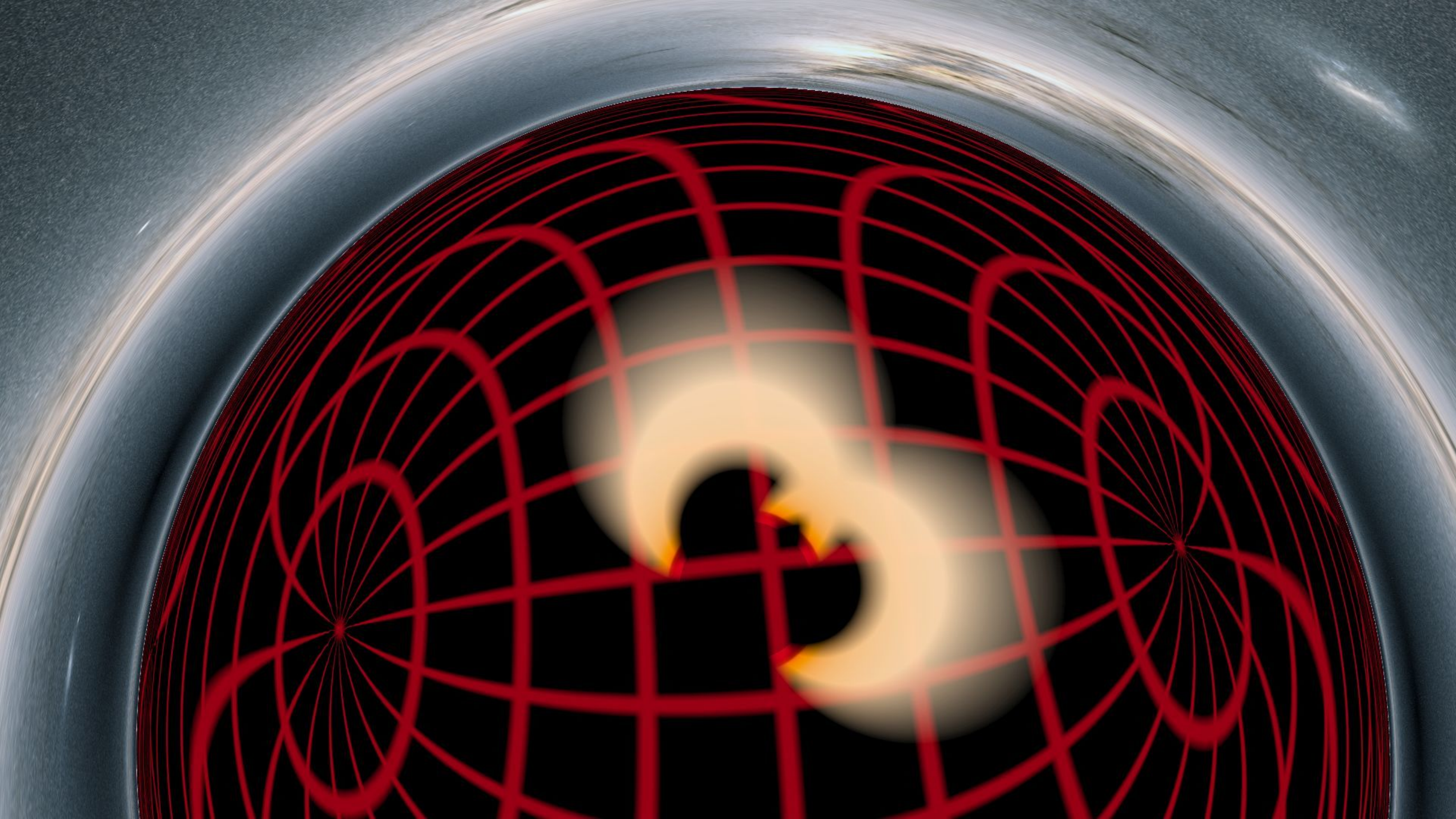}
    \includegraphics[scale=.125]{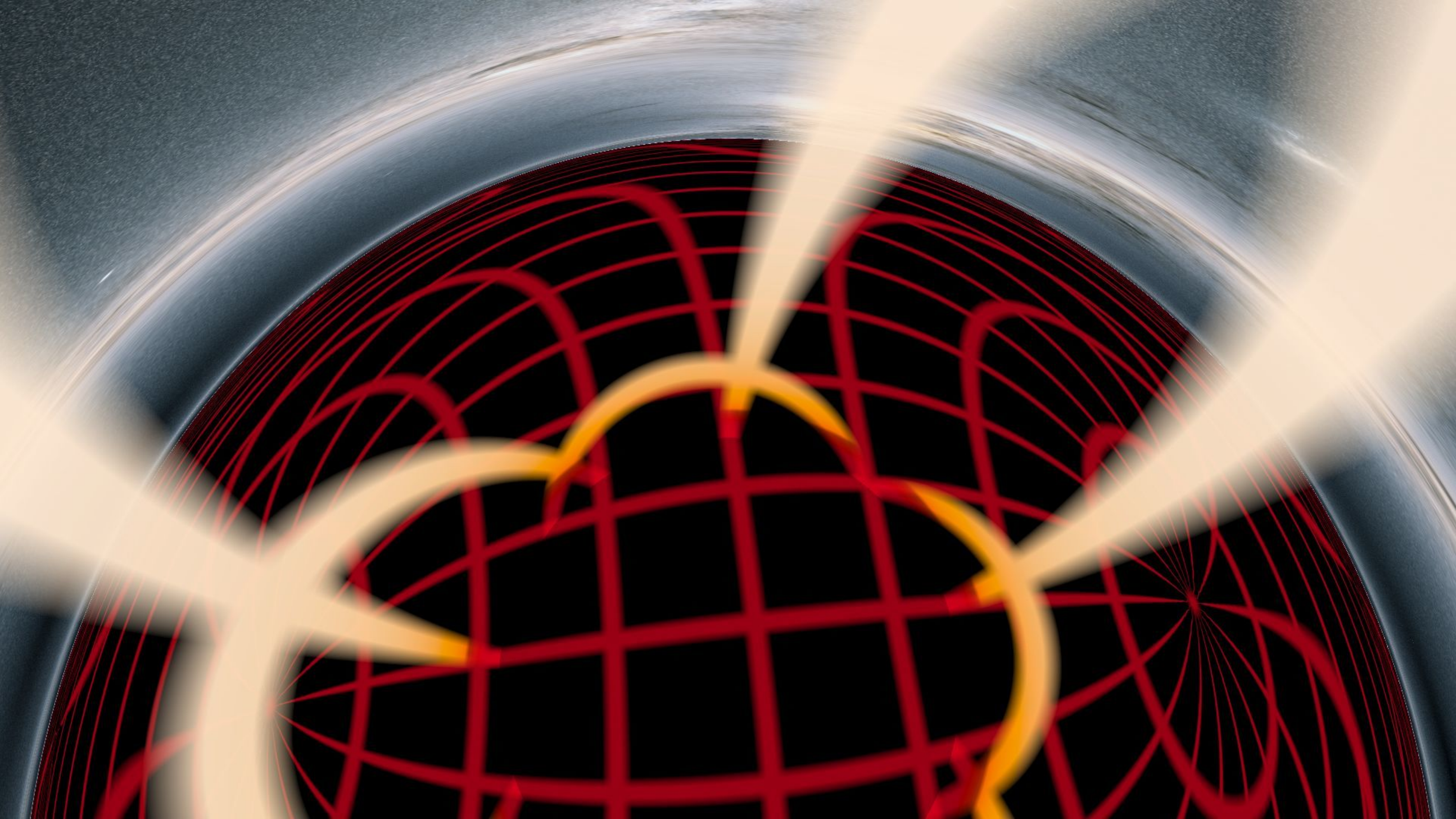}
    \includegraphics[scale=.125]{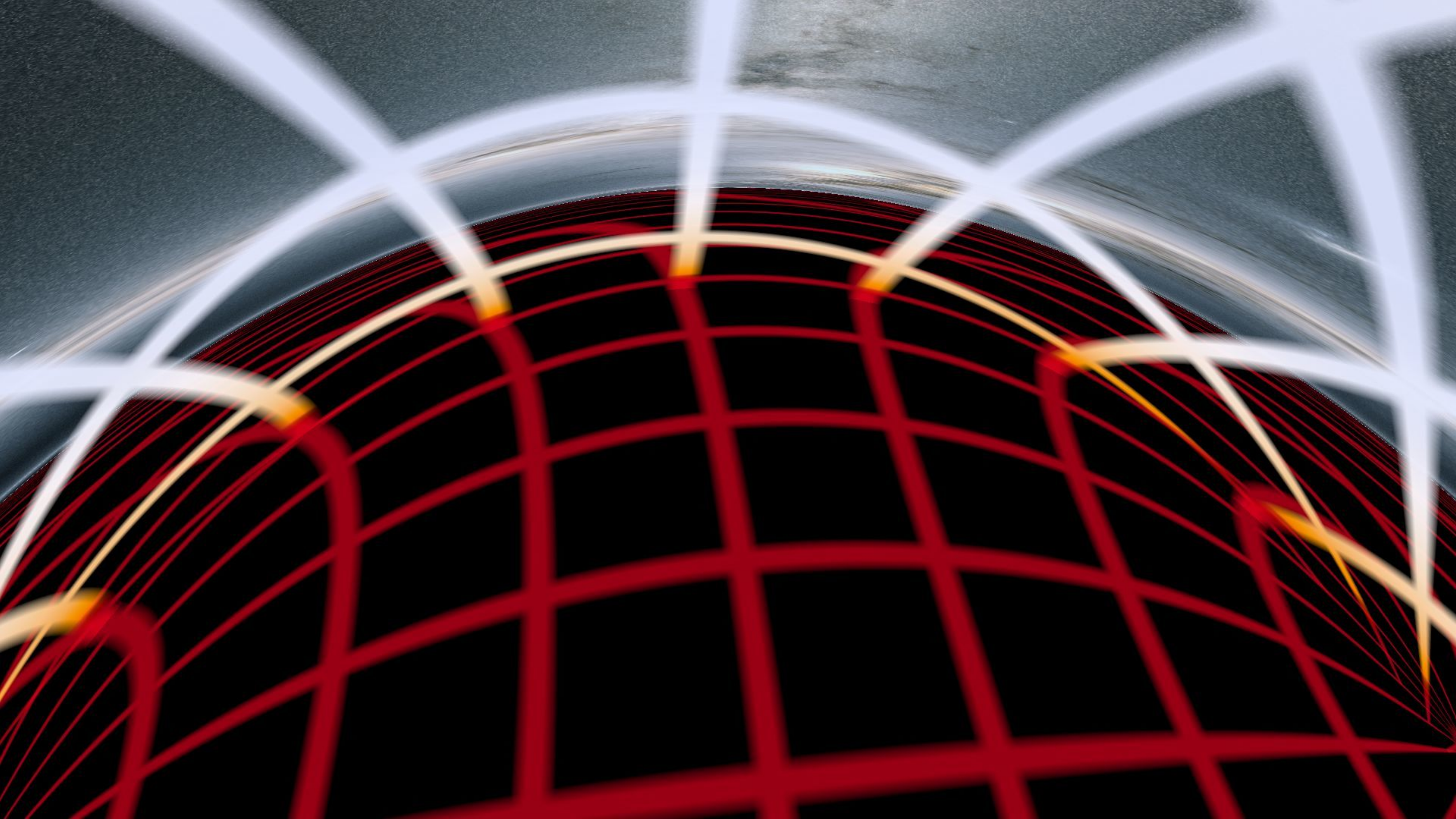}
    \includegraphics[scale=.125]{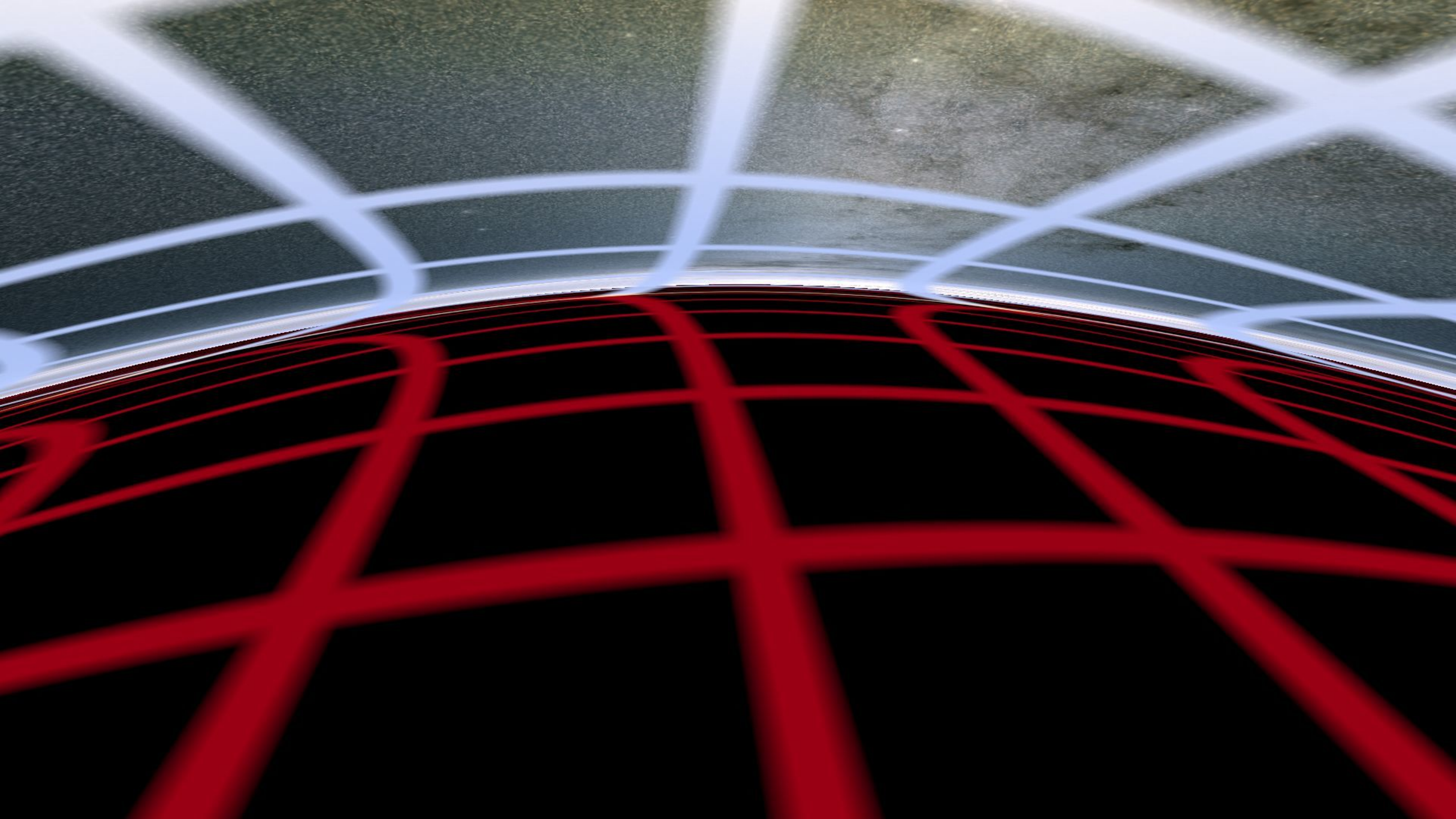}
    \includegraphics[scale=.125]{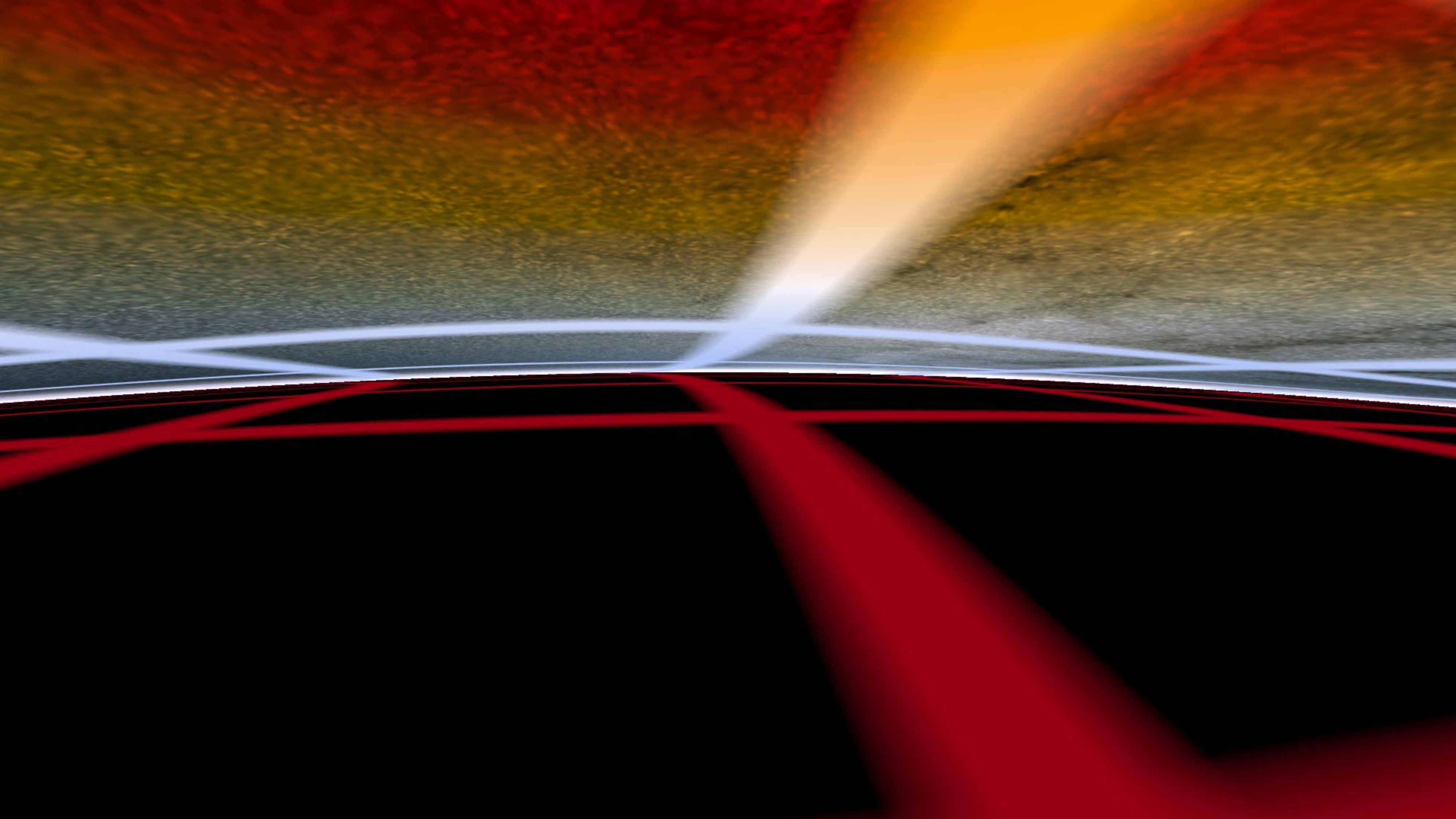}
    \includegraphics[scale=.125]{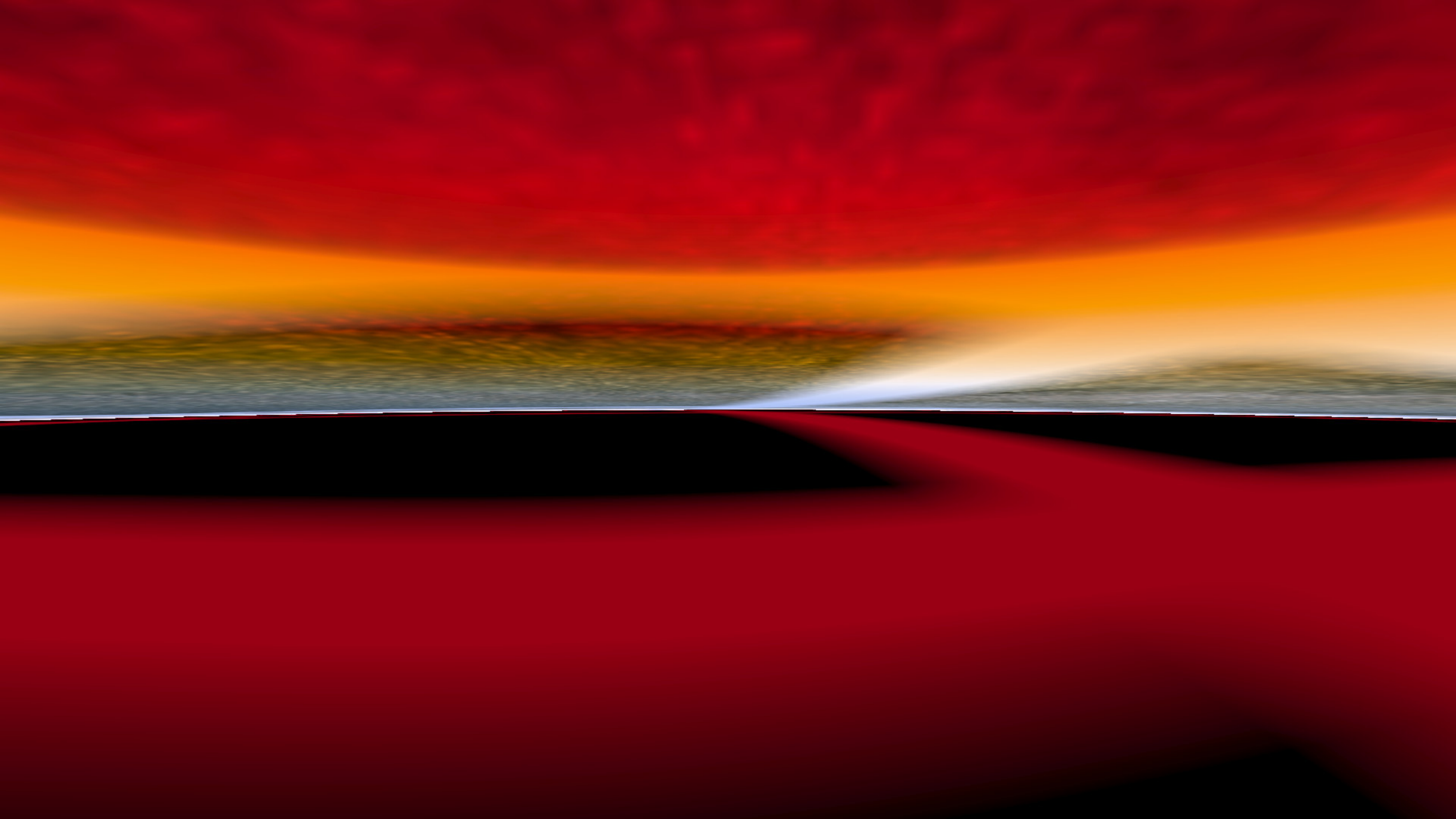}
    \caption[Visualization of falling into a Schwarzschild black hole]{
    \label{schwviz}
Eight frames from a visualization of the view
seen by an observer who free-falls radially
through the horizon of a Schwarzschild black hole
\cite{Hamilton:2010my}.
The infaller is on a geodesic that starts at rest at infinity
and falls radially with zero angular momentum.
From left to right and top to bottom,
the observer is at radii
%$3$,
$1.5$,
$1.01$,
$0.99$,
$0.9$,
%$0.8$,
$0.5$,
$0.1$,
$0.01$,
and
$0.001$
horizon radii.
The dark red grid is the black hole's past horizon,
which in a real black hole is replaced by the dimming, redshifting
surface of the star or whatever else collapsed to the black hole long ago.
The event or future horizon is painted with a grid
colored with an appropriately red- or blue-shifted blackbody color.
The background is an image of the Milky Way from Gaia Data Release~3
\cite{GaiaDR3:2020}.
    }
    \end{center}
    \end{figure*}
}

%--------------------
% FIG
\newcommand{\kerrvisiblefig}{
    \begin{figure}[t!]
    \begin{center}
    \leavevmode
    \includegraphics[scale=.7]{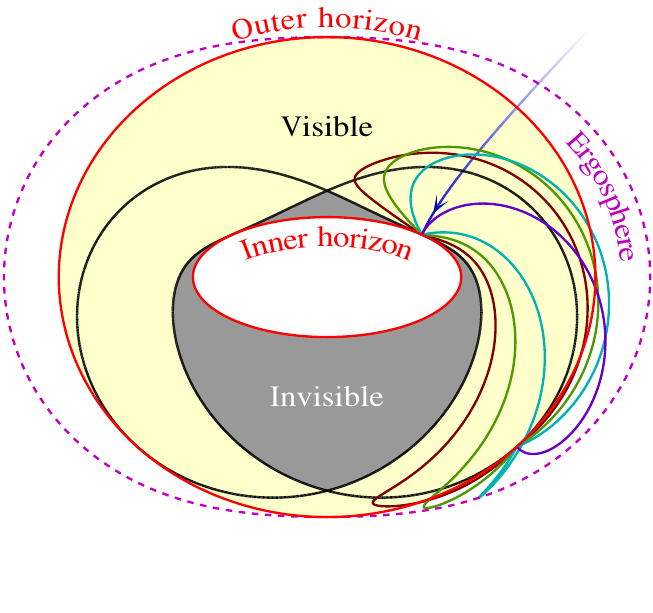}
    \caption[Regions visible and invisible to an infaller at the inner horizon]{
    \label{kerrvisible}
Regions between the outer and inner horizons visible and invisible
to an infaller who reaches the inner horizon of a Kerr black hole.
The black hole here has spin parameter $a = 0.8 \Mbh$,
and the infaller (blue line) falls with zero angular momentum $L = 0$
along a geodesic at latitude $\theta_0 = 45^\circ$.
The boundary of the (in)visible region is set by null trajectories with
infinite Carter constant $\KCarter$
and asymptotically small azimuthal angular momenta $J$;
such trajectories, dragged around by the rotation of the black hole,
spiral around the black hole divergingly many times,
hugging the $J = 0$ surface (black lines)
all the way between the outer and inner horizons.
Null geodesics with infinite $\KCarter$
and various azimuthal angular momenta $J$ are shown
on the radius-latitude $r$-$\theta$ plane
(projected to azimuth $\phi = 0$)
as solid lines passing through the observer at the inner horizon.
To reach the observer at latitude $\theta_0$,
$|J|$ cannot exceed $J_{\max} = \sin\theta_0$;
the trajectories shown have
$|J|/J_{\max}$ equal to
0 (black),
$\tfrac{1}{4}$ (brown),
$\tfrac{1}{2}$ (green),
$\tfrac{3}{4}$ (cyan),
and
1 (purple).
For each $|J| < J_{\max}$ there are two trajectories,
which approach the observer from respectively lower and higher latitude.
Null geodesics with infinite $\KCarter$ start as outgoing at
(infinitesimally outside) the outer horizon,
switch to ingoing inside the ergosphere,
fall back through the outer horizon, down to the inner horizon.
All the geodesics start on the outer horizon at the same latitude antipodal,
$\theta = \tfrac{1}{2}\pi - \theta_0$,
to the final latitude $\theta_0$ at the inner horizon
(but not antipodal in azimuth).
Compare to Figure~\ref{schwJoovis} for a Schwarzschild black hole.
    }
    \end{center}
    \end{figure}
}

%--------------------
% TAB
\newcommand{\Ptab}{
    \begin{table}[bt!]
    \caption[Signs of $P_t$ and $P_x$ in various regions of the Kerr-Newman geometry]{
Signs of $P_t$ and $P_x$ in various regions of the Kerr-Newman geometry
    }
    \begin{center}
    \label{Ptab}
    \begin{tabular}{lr}
\hline
\hline
Region & Sign \\
\hline
Universe, Wormhole, Antiverse & $P_t < 0$ \\
Parallel Universe, Parallel Wormhole, Parallel Antiverse & $P_t > 0$ \\
Black Hole & $P_x < 0$ \\
White Hole & $P_x > 0$ \\
Horizon, Inner Horizon & $P_t = P_x < 0$ \\
Parallel Horizon, Parallel Inner Horizon & $- P_t = P_x < 0$ \\
Antihorizon, Inner Antihorizon & $- P_t = P_x > 0$ \\
Parallel Antihorizon, Parallel Inner Antihorizon & $P_t = P_x > 0$ \\
\hline
    \end{tabular}
    \end{center}
    \end{table}
}

\date{\today}

\begin{abstract}
It is widely repeated in the popular literature and elsewhere
that the singularity at the center of a black hole is a point.
It is not true.
Two observers who free-fall into a spherical black hole
along two different angular trajectories at the same time $t$
do not encounter each other at the central singularity; rather,
they lose causal contact with each other already well away from the singularity.
Counterintuitively,
in general relativity two points can be spatially close yet causally distant.
The singularity is a surface, not a point.
The story for rotating black holes is more complicated,
but the same conclusion holds.
For a rotating black hole, the singular surface almost certainly resides
at its inner horizon, where even the tiniest classical or quantum perturbations
ignite the exponential mass inflation instability,
precipitating collapse to a spacelike singular surface.
There are implications for quantum gravity.
We argue that, whatever the ultimate theory of quantum gravity may be,
the quantum states of a black hole probably reside
at its effectively 2-dimensional singular surface,
which coevolves unitarily with, and in thermodynamic equilibrium with,
the hot atmosphere of trapped Hawking radiation that the black hole generates
within its event horizon.
%Whereas in general relativity time comes to an end at the singular surface,
%quantum mechanics requires that the singular surface must continue
%to evolve unitarily.
%%the quantum mechanical equivalent of classical determinism.
%We argue that, whatever the ultimate theory of quantum gravity may be,
%the most likely picture is that the singularity is a hot
%effectively 2-dimensional surface coevolving unitarily with,
%and in thermodynamic equilibrium with,
%its attendant Hawking radiation.
%%The states of a black hole reside at or near its singular surface,
%%not its horizon.
\end{abstract}

\maketitle

\section{Introduction}
\label{intro-sec}

%https://bigthink.com/starts-with-a-bang/singularities-dont-exist-roy-kerr/
%http://www.madore.org/~david/math/kerr.html

The most energetic places in the known Universe
occur at the Big Bang and at the singularities of black holes.
Unsurprisingly,
not only seasoned experts
but also students and the general public are fascinated by these places,
where the known laws of physics break down.

The purpose of this paper is to push back against the widespread misconception
that the singularity at the center of a black hole is a point.
Sections~\ref{introschw-sec} and~\ref{introkerr-sec} of this Introduction
offer context reviewing mainstream treatments of singularities within spherical
(Schwarzschild) and rotating (Kerr) black holes, respectively.

Section~\ref{schw-sec} demonstrates a central result of this paper,
that the singularity at the center of a
spherical, Schwarzschild black hole is best characterized as a surface, not a point.
The key argument is that different angular locations on the singularity
are causally disconnected, in spite of being metrically close.
%The singularity of a Schwarzschild black hole,
%where general relativity breaks down,
%is the boundary of a 3+1-dimensional spacetime,
%asymptoting to a 3-dimensional spacelike surface.
%Extended in the time direction, the asymptotic boundary to
%the Schwarzschild singularity is a 3-dimensional spacelike surface.
%Support for this claim comes primarily from the causal and geodesic structure of the spacetime.
Section~\ref{infallers-sec} calculates the boundary of the region
of visibility seen by an infaller approaching the singularity,
who loses all causal contact with neighboring infallers approaching the same singularity.
The loss of causal contact between neighboring regions is a known feature
of gravitational collapse in both black holes and cosmology
\cite{Belinskii:1970,Belinskii:1982,Andersson:2005}.

Geroch, Kronheimer, and Penrose \cite{Geroch:1972}
proposed extending a general relativistic spacetime
by including on its boundary ``ideal points,''
defined to be distinct if they do not share the same past or future,
that is, if they are causally disconnected.
Ideal points include not only those on a singularity,
but also points at infinity.
Reference~\cite{Geroch:1972} emphasize that the definition of ideal points
depends on the causal structure of a spacetime, not on its metric.
%The widespread use of Penrose diagrams \cite{Penrose:1964ge}
%attests to the acceptance of \cite{Geroch:1972}'s ideas.
%\cite{Marolf:2003} addresses some subtleties that arise when
%dealing with more general manifolds.
The review by \cite{GarciaParrado:2005} offers an entry to the literature
on causal boundaries.

%The singular surface advocated in the present paper should be understood
%in the same sense as \cite{Geroch:1972}.
%By no means do we suggest that the singular surface
%is literally a 2-dimensional object in a final quantum theory of gravity.

%In the context of cosmology,
%the disconnection of causal regions near the singularity of a collapsing
%Friedmann-Lema{\^i}tre-Robertson-Walker universe is known as
%``asymptotic silence'' \cite{Andersson:2005}.
%As distinct observers asymptotically approach the singularity from different
%angles, their past light cones funnel onto their own past worldlines,
%leading to causally disconnected endpoints, even though a na\"ive extrapolation
%would imply that the angular dimensions at the singularity are topologically
%degenerate.

Section~\ref{kerr-sec} generalizes to a rotating, Kerr black hole.
Unlike the Schwarzschild singularity,
the singularity within a mathematical Kerr black hole is often described as
a timelike, one-dimensional ring.
However, as argued in \S\ref{massinflation-sec},
in a real rotating black hole,
the geometry of a Kerr black hole is almost certainly
truncated at its inner horizon
by classical or quantum perturbations
that trigger the mass inflation instability,
precipitating collapse to a spacelike singularity.
If so, the singularity of a Kerr black hole resides at its inner horizon,
and its structure mirrors the spherical analysis of
\S\ref{schw-sec}.

%Section~\ref{geom-sec} argues that the singular surface can be assigned
%an intrinsic observer-independent geometry.
%For observers inside as well as outside the event horizon,
%the perceived boundary of spacetime is set by the past horizon,
%and the past horizon possesses a 2D geometry whose area determines
%the entropy of the black hole.
%%In both Schwarzschild and Kerr cases,
%%the singularity is a spacelike surface with the topology of a 2-sphere,
%%the points of which are causally separated from each other.

Section~\ref{qg-sec} discusses some implications for quantum gravity.
%We caution that by no means do we suggest that the singularity
%remains a 2D surface in quantum gravity.
%In string theory and loop quantum gravity for example,
%general relativity emerges only as an effective large scale description.

For reference,
Appendices~\ref{schwarzschildgeodesics-sec}
and~\ref{kngeodesics-sec}
review the well-known solutions of the geodesic equations
in respectively Schwarzschild and Kerr black holes.

\subsection{Spherical black hole review}
\label{introschw-sec}

Most textbooks on general relativity tend to be guarded in their treatment of
singularities of black holes,
even in the simplest case,
that of a spherical uncharged black hole,
a Schwarzschild black hole.
Commonly the textbooks point out that the tidal force diverges
at the singularity of a Schwarzschild black hole at zero radius $r = 0$,
and that the usual Schwarzshild time and radial coordinates $t$ and $r$,
which are respectively timelike and spacelike outside the horizon,
switch to become respectively spacelike and timelike inside the horizon.
An infaller who falls inside the horizon is inexorably dragged inward
toward the singularity.
Absent a consensus theory of quantum gravity,
what happens at the singularity remains enigmatic.

One classic text that does address the Schwarzschild singularity is that of
Rindler \cite{Rindler:1977}, who rightfully disputes the notion of a singular point but nonetheless mischaracterizes the singularity as a 1-dimensional line. He writes on page 159,
``we tend to think of the Schwarzschild locus $R = 0$ [$\dots$]
as a single spatial point, the permanent center of the horizon sphere [$\dots$]
And this is false!''
Rindler then clarifies on page 161
while describing the Kruskal \cite{Kruskal:1960,Szekeres:1960} diagram
of a Schwarzschild black hole:
``the universe suddenly appears out of nowhere as an infinite line''
(a white hole singularity), which
``immediately flares into a long drawn out double trumpet''
before collapsing back to a line
(a black hole singularity)
``momentarily, and then again nothing.''
The singular lines are spacelike,
consistent with the conceptual picture that proper time comes into existence
at the white hole singularity,
and comes to an end at the black hole singularity, after the throat of the so-called Einstein-Rosen bridge becomes infinitely long and vanishingly thin.

A Kruskal diagram is a spacetime diagram
constructed from radial and time coordinates arranged
such that radially moving light always moves at 45$^\circ$.
Its virtue is that it clarifies the causal structure of the geometry,
since timelike worldlines must evolve generally upward,
at less than $45^\circ$ from vertical.
A Penrose diagram \cite{Penrose:1964ge}, Fig.~\ref{schwpenrose},
shares with a Kruskal diagram the property that radially moving light
always moves at 45$^\circ$,
but in addition maps spatial and temporal infinity to a finite position.
Like the Kruskal diagram,
the Penrose diagram shows that the past (white hole) and future (black hole)
singularities are extended in the spacelike $t$ direction.

Rindler's \cite{Rindler:1977} characterization of the singularities as lines
stems from the fact that they occur at zero radius $r = 0$,
where the angular dimensions shrink to zero.
A primary purpose of the present paper is to show that this seemingly
obvious inference is false.
The root reason is that in general relativity,
being spatially close does not necessarily imply being causally close.

A sentiment similar to Rindler's is expressed in the weighty tome by
Chandrasekhar \cite{Chandrasekhar:1983}, who,
in discussing the Kruskal diagram of the Schwarzschild geometry,
writes on page 92,
``the centre emerges as a singular line of the Schwarzschild space-time.''

\schwpenrosefig

%https://www.space.com/what-happens-black-hole-center

The popular literature tends to be more pointed.
It is unified in its declaration
that the singularity of a black hole is
%0-dimensional.
a point.
Brian Greene in ``Fabric of the Cosmos'' \cite{Greene:2004}
writes on page 337:
\begin{quoting}
According to general relativity, all that makes up a black hole
is crushed together at a single minuscule point at the black hole's center.
\end{quoting}
%Greene in ``The Hidden Reality'' \cite{Greene:2011}
%offers an optimistic view about string theory's ability to solve singularities,
%but admits that the real singularities of black holes
%are not solved by string theory.
Neil deGrasse Tyson in ``Death by Black Hole'' \cite{Tyson:2007}
writes on page 284:
\begin{quoting}
Meanwhile the stuff within the event horizon has collapsed to
an infinitesimal point at the black hole's center.
\end{quoting}
Marcia Bartusiak in the preface of ``Black Hole'' \cite{Bartusiak:2016}
writes:
\begin{quoting}
The idea of the black hole is actually quite simple.
It has a mass, and it has a spin.
In some ways, it's as elementary an entity as an electron or a quark.
But what confounded physicists for so long was the black hole's ultimate nature:
it's matter squeezed to a point.
\end{quoting}

The picture that the singularity is a point is seemingly well-founded.
It comes from the fact that
a spherical black hole is described by
the Schwarzschild \cite{Schwarzschild:1916a} metric
(eq.~(\ref{schwmetric}) in Appendix~\ref{schwarzschildgeodesics-sec}),
and that the central singularity, where the curvature diverges,
is at zero radius,
\mbox{$r = 0$}.
The radius of a 2-dimensional sphere enclosing the singularity
diminishes to zero at the singularity.

A mathematical relativist would be more cautious in their language.
They would point out that the central ``point'' of the Schwarzschild metric
is not part of the spacetime manifold, because general relativity fails there:
there is no locally inertial frame at the singularity.
Rigorously, within general relativity,
it is legitimate to make statements only about regions asymptotically close to,
but not at, the singularity.
The present paper will follow the relativist's caution.

%In its article on black branes,
%wikipedia
%\url{https://en.wikipedia.org/wiki/Black_brane}
%states:
%\begin{quote}
%``However, many physicists tend to define a black brane separate
%from a black hole, making the distinction that the singularity of a black brane
%is not a point like a black hole, but instead a higher dimensional object.''
%\end{quote}
%I too am guilty of repeating the popular wisdom.
%In the ``Falling into a Black Hole'' website\footnote{
%\url{http://jila.colorado.edu/~ajsh/bh/singularity.html}}
%I constructed in 1997, I wrote:
%\begin{quote}
%``The singularity: an infinitely convoluted point of infinite curvature.''
%\end{quote}

\subsection{Rotating black hole review}
\label{introkerr-sec}

The Schwarzschild geometry describes a spherical, nonrotating black hole.
But real black holes rotate \cite{Reynolds:2021}.
A celebrated example is Cygnus-X1,
an X-ray binary consisting of a $19 \unit{\Msun}$ giant star
accreting onto a $15 \unit{\Msun}$ black hole
\citep{Orosz:2011},
whose spin parameter $a$ has been measured to be at least 0.95
of the maximum possible spin
\citep{Gou:2011nq}.
%The exceptionally well-measured gravitational wave event GW250114
%is interpreted as the merger of two $\approx 33 \unit{Msun}$ black holes
%with small initial spins $\lesssim 0.26$,
%and a final spin of $0.68 \pm 0.01$ of the maximum possible
%\cite{Abac:2025}.

%The story with rotating black holes is more complicated than that
%of spherical black holes,
%but the outcome is probably the same:
%the singularity of a rotating black hole is a surface,
%with spherical topology like that of a spherical black hole.

Rotating black holes are
%usually
described by the Kerr \cite{Kerr:1963,Kerr:2009}
geometry.
The big difference between a spherical (Schwarzschild) and a rotating (Kerr)
black hole is that a rotating black hole has not one but two horizons,
an outer and an inner horizon.
As in a Schwarzschild black hole,
space falls faster than light inside the outer horizon of a Kerr black hole
\citep{Hamilton:2004au}.
But in a Kerr black hole,
%the centrifugal repulsion from the black hole's spin grows more rapidly inward
%than the gravitational attraction from the black hole's mass.
centrifugal repulsion slows the inflow of space inside the outer horizon,
back down to the speed of light at the inner horizon.
%The centrifugal force causes the horizons to bulge into spheroids.
%The outer and inner horizons form confocal spheroids
%with a ring singularity at their focus.

As first elucidated by Carter in 1968 \cite{Carter:1968a},
in the exact, analytically extended,
mathematical solution for the Kerr geometry,
the inner horizon is a gateway to remarkable and astonishing phenomena,
including wormholes, white holes, and
%, most notable for the present discussion,
a central ring singularity.
The ring singularity is timelike and, thanks to the centrifugal force,
infinitely gravitationally repulsive.
Its ring-like nature can be ascertained by noting that it occurs
along the equatorial plane ($\theta = \pi/2$) at 
Boyer-Lindquist \cite{Boyer:1967} radius $r=0$,
where $r$ is an oblate spheroidal coordinate related to Cartesian coordinates
$\{x,y,z\}$ by the transformation
$x^2+y^2+z^2=r^2+a^2\sin^2\theta$.
Whenever the spin $a$ is nonzero, the singularity spreads out 
into a ring of radius $a$.

The divergingly repulsive character of the ring singularity
is not its only unphysical feature.
Adjacent to the ring singularity is an azimuthal tunnel through
which an observer, if moving sufficiently rapidly retrograde,
can go into their own past, violating causality \cite{Carter:1968a}.
The ring singularity is a source of unpredictability:
timelike and lightlike worldlines emerge from it without causal precedent.
Naturally one should be skeptical that
timelike singularities, violations of causality, wormholes,
and other such phantasmagoria could happen in reality.

Mathematicians call the boundary of unpredictability the Cauchy horizon.
In a Kerr black hole, the Cauchy horizon is the inner horizon.
Section~\ref{massinflation-sec} argues that in a realistic rotating black hole,
the inner horizon replaces the ring singularity as the boundary
of the spacetime.

\penrosemassinflationfig

The analysis of the nature of the Kerr interior in \S\ref{kerr-sec}
will be focused on its overall causal structure and the behavior
of ingoing and outgoing geodesics.
The Kerr geometry's inner horizon is in fact composed of two distinct surfaces,
here called the outgoing inner horizon (through which outgoing geodesics pass)
and the ingoing inner horizon (through which ingoing geodesics pass).
Qualitatively, outgoing means going against the superluminal inflow
of space between the horizons,
while ingoing means going with the flow.
Of course both outgoing and ingoing geodesics are compelled to fall inward,
to smaller radius $r$, by the superluminal inflow.

The causal structure of the Kerr geometry is illuminated by its Penrose diagram,
Figure~\ref{penrosemassinflation}.
%A Penrose diagram
%\cite{Penrose:1964ge}
%is a spacetime diagram drawn so that
%(a) radially outgoing and ingoing light rays move at $45^\circ$ from vertical,
%and (b) points at infinity (past, future, or spatial)
%are brought to finite position.
%A Penrose diagram brings out the causal structure because
%all worldlines of actual massive objects must move at less than the speed
%of light, hence generally upward, at less than $45^\circ$ from vertical,
%in the diagram.
%Of course, a Penrose diagram does not capture all the richness of
%a spacetime such as Kerr,
%because the diagram retains only one of the three spacelike directions.
%
As illustrated in the diagram,
outgoing geodesics fall through the outgoing inner horizon
into a region of the Kerr geometry where the time coordinate $t$
(the coordinate with respect to which
the geometry is time-translation symmetric)
progresses backward;
while ingoing geodesics fall through the ingoing inner horizon to a region
where the time coordinate $t$ progresses forward.
The two regions beyond the inner horizons can be called
outgoing and ingoing wormholes,
because the centrifugal force continues to slow the inflow of space,
turning it around inside the wormholes to become an outflow of space.

The Kerr geometry can be analytically continued beyond the wormholes
to a white hole and other universes.
However, that story will not be pursued here, because none of it happens in reality.
Instead, as will be elaborated in \S\ref{kerr-sec},
the inner horizon will itself probably become a spacelike singular surface,
rendering the rotating black hole's interior and singularity
topologically identical to that of Schwarzschild.

\section{Spherical black holes}
\label{schw-sec}

\schwJoovisfig

According to the textbook claims of \S\ref{introschw-sec},
the central singularity in a Schwarzschild black hole has zero radius
and therefore should be viewed as asymptotically approaching a point or line
with vanishing angular dimensions.
The goal of this section is to show that this view is misleading,
not only from the perspective of causal geodesics
(\S\ref{infallers-sec} and \S\ref{contra-sec})
but also from ray-traced visualizations tracking observables within orthonormal frames (\S\ref{view-sec}).

\subsection{Two infallers do not meet each other at the singularity}
\label{infallers-sec}

Fig.~\ref{schwJoovis}
shows the trajectories of two observers, one blue, one purple, who free-fall
radially into a Schwarzschild black hole along two different angular positions.
You can imagine that they do their best to fall in at the same time $t$,
although because of the finite light travel time between them,
they always see the other observer farther out in radius
than they themselves are.

The defining feature of a black hole is that space inside the horizon falls
faster than light \citep{Hamilton:2004au}
(the special relativistic rule
is that nothing can move {\em through\/} space faster than light;
spacetime itself can do whatever general relativity prescribes).
So light that is emitted from inside the horizon is forced to move inward,
even if the light is trying to move outward.
%If the light is trying to move outward,
%the best it can do is a Michael Jackson moonwalk inward.

Consider what the blue infaller sees as they approach the singularity.
As Fig.~\ref{schwJoovis} (which is mathematically accurately drawn)
illustrates, the blue infaller near the singularity cannot see all
of the spacetime inside the horizon.
There is a certain maximum angular motion that a light ray can have
inside the black hole's horizon.
That maximally sideways-moving light bounds what an infaller can see.
The region beyond that boundary is invisible to the infaller.

The calculation of geodesics in the Schwarzschild geometry is
outlined in Appendix~\ref{schwgeodesics-sec}.
%is a calculation that every aspiring student of general relativity should do.
The null geodesics with the maximum angular motion inside the horizon
turn out to be those with infinite angular momentum per unit energy.
The expression for the radius $r$ versus angle $\theta$
along such geodesics (eq.~(\ref{thetaooorbit})),
as seen by an observer at angular position $\theta_\obs$,
is pretty,
\begin{equation}
\label{rschwJoo}
  r
  =
  r_s
  \sin^2\!\left( {\theta - \theta_\obs \over 2} \right)
  \ ,
\end{equation}
where $r_s$ is the Schwarzschild radius, the radius of the horizon.
Equation~(\ref{rschwJoo}) is the equation of a cardioid.
The cardioid is the heart-shaped boundary of the zone of invisibility
depicted in Fig.~\ref{schwJoovis}.
The cardioid just kisses the horizon at the antipode
of the observer's angular position $\theta_\obs$.

The last that the blue infaller sees of the purple infaller
is at the boundary of the invisible gray region,
already well away from the singularity.
Far from encountering each other at the singularity,
the blue and purple infallers lose causal contact with each other
already some distance from the singularity.
%For the two infallers to bump fists at the singularity,
%they would have to extend their fists sideways faster than the speed of light.
%Put another way,
%the diverging tidal force near the singularity funnels infallers
%along ever narrowing radial trajectories.
Because both infallers fall at the same time $t$,
both infallers would appear to follow the same
worldline on the Penrose diagram~\ref{schwpenrose}.
The singularity is therefore causally separated
not only along the spacelike $t$ direction,
but also along angular directions for two infallers with the same
coordinate $t$.
Since the two infallers fall to causally disconnected places,
the singularity cannot be a point or a line: it must be a surface.

\schwvizfig

\subsection{Contradiction?}
\label{contra-sec}

The argument seems to have led to a contradiction.
On the one hand, two infallers reaching the singularity are causally separated.
On the other hand, the Schwarzschild metric shows that the spatial distance
between the infallers goes to zero at the singularity.
The resolution of this conundrum is that, in general relativity,
two points can be spatially close, yet causally distant.

In everyday experience,
things that are spatially close are causally close.
It can be hard to adjust one's intuition to the fact
that this is not true in general relativity.
But being close in one sense
does not necessarily imply being close in another sense.
For example,
in special relativity the spacetime distance $\dd s$
along a null (lightlike) path is zero,
even though the null path connects points
that are separated in both time and space.
Relativity has a tendency to defy intuition.

Where does the intuition that spatially close implies causally close go wrong?
In the Schwarzschild geometry,
the spatial separation between infallers approaching the singularity
is measured along the angular direction.
But, as Fig.~\ref{schwJoovis} illustrates,
the angular region that an infaller can measure with a ruler
diminishes near the singularity to a sliver,
shrinking to a cusp at the singularity.
Near the singularity,
there is no actual observer who can measure anything but a small part
of the circumference of a sphere that encloses the singularity.

%The inference that the circumference of the diminishing sphere
%goes to zero at the singularity is based on an extrapolation.
%The diverging tidal force stretches in the radial direction
%but collapsing in the angular direction.

The shortest causal path between two infallers
reaching the singular surface at the same time $t$ but at two
different angular positions
is a pair of null geodesics each with the maximum possible angular momentum,
as illustrated by the red lines in the right panel
of Figure~\ref{schwJoovis}.
A quantitative measure of the causal distance along null geodesics
in general relativity is the affine distance
\cite{Hamilton:2010my}.
Appendix~\ref{affine-sec} defines affine distance,
and shows that, up to an arbitrary overall normalization factor,
the shortest causal distance between infallers separated by angle $\theta$
is given by equation~(\ref{lambdashortestcausal}).

\subsection{The view near the singularity}
\label{view-sec}

The notion that the singularity is a surface becomes more compelling
when one considers what an infaller sees as they approach the singularity.
Because space falls faster than light inside the horizon,
an observer inside the horizon of a Schwarzschild black hole
is always looking at light that was emitted above them,
even if they are looking downward.
For example,
if the observer falls feet first through the horizon,
they continue to see their feet below them,
but they are seeing an image of their feet as they used to be above their head.
An infaller looking down never gets to see the singularity below them.

Fig.~\ref{schwviz}
shows eight frames from a general-relativistically ray-traced visualization
of what an observer would see free-falling into a Schwarzschild black hole,
an issue first addressed by \cite{Hamilton:2010my}.
In the visualization,
%in Fig.~\ref{schwviz},
the observer starts from rest at infinity,
and free-falls radially with zero angular momentum.
%(constants of motion are $E = 1$ and $L = 0$).

The black hole in Fig.~\ref{schwviz} is painted with a dark red grid,
the past horizon of the black hole \cite{Hawking:1973}.
In a real black hole,
the past horizon is physically the redshifting, dimming surface
of the star or whatever else collapsed to the black hole long ago.
The infaller does not catch up with the image of the collapsed star;
rather, the image of the collapsed star remains ahead of the infaller,
still redshifting and dimming away.

When the infaller falls through the horizon,
the horizon appears to split into two.
In the Penrose diagram~\ref{schwpenrose},
an observer passing through the Horizon goes from seeing
the surface labeled Antihorizon
to seeing two surfaces, the Parallel Horizon and the Horizon.
The event horizon, or future horizon \cite{Hawking:1973},
the whitish grid in Fig.~\ref{schwviz},
appears out of nowhere.
This is the surface of no return,
within which space falls faster than light.
The event horizon becomes visible only after the infaller has passed through it.
The event horizon, and the sky above, are colored in Fig.~\ref{schwviz}
with an appropriately red- or blue-shifted blackbody color.

As the infaller approaches the singularity,
they have the impression of landing on a flat plane.
The two horizons, the event and past horizons,
appear to converge toward this flat plane.
The infaller has the impression that they finally catch up with
the past horizon at the singularity.
The appearance of flattening can be attributed to the diverging tidal force
near the singular surface.
The mathematical details that give rise to the observed surface-like behavior
are given in Appendix~\ref{flattening-sec}.

\section{Rotating black holes}
\label{kerr-sec}

%Unlike the singularity of the Schwarzschild geometry, the singularity in mathematical models of black holes with two horizons (like the Kerr geometry) is timelike and no longer has the properties of a 2-dimensional surface.
In rotating black holes, the singularity is commonly presented as a ring of
zero thickness, hidden in a timelike region behind two horizons.
However, as claimed in \S\ref{introkerr-sec},
in realistic settings the Kerr geometry's inner horizon is unstable
and should collapse into a spacelike singular surface,
blocking off any hypothetical wormholes and ring singularities.
Section~\ref{massinflation-sec} elaborates this argument.
Then, \S\ref{kerrgeodesics-sec} proceeds to show how two infallers
will not be causally close upon reaching the inner horizon
(and therefore the resulting singularity),
mirroring the Schwarzschild analysis of \S\ref{infallers-sec}.

\subsection{Mass inflation}
\label{massinflation-sec}

In the same year (1968) that Carter \cite{Carter:1968a}
revealed the Kerr geometry in all its glory,
Penrose \cite{Penrose:1968} pointed out
that an observer who reaches the outgoing inner horizon of a black hole
will see the outside Universe infinitely blueshifted
(Penrose considered a spherical charged, Reissner-Nordstr\"om, black hole,
but the same holds true for a rotating, Kerr, black hole).
In the Penrose diagram~\ref{penrosemassinflation},
the outgoing inner horizon is a null line extending from
the future infinity of the Universe.
Penrose conjectured that the diverging concentration of energy collecting
at the inner horizon would cause it to collapse,
preventing any of the bizarre features of the geometry inside the inner horizon
from coming into existence.

The reason for the infinite blueshift at the inner horizon
has to do with the behavior of geodesics near horizons.
As first pointed out by Carter \cite{Carter:1968c},
the Hamilton-Jacobi equation in the Kerr geometry is separable,
and therefore geodesics are exactly solvable,
as outlined in Appendix~\ref{kngeodesics-sec}.
It is convenient to describe Kerr geodesics in terms of a set of
Hamilton-Jacobi parameters $P_t$, $P_x$, $P_y$, and $P_\phi$,
defined by equations~(\ref{Ptphixy}).
%In the Kerr geometry, outgoing and ingoing geodesics are distinguished quantitatively by the sign, positive or negative, of the Hamilton-Jacobi parameter $P_t$ defined by equation~(\ref{Ptphi}).
These parameters depend on the conserved energy $E$
and azimuthal angular momentum $L$ of the geodesic,
but they themselves are not constants of motion.
%Prograde geodesics, those circulating with the spin of the black hole,
%experience extra centrifugal force.
%Although a prograde geodesic must necessarily be ingoing at the outer horizon,
%the extra centrifugal force can turn it around between the horizons
%so that it becomes outgoing at the inner horizon.
%Retrograde geodesics experience diminished centrifugal force,
%and remain ingoing at the inner horizon.
As the Hamilton-Jacobi parameters change along a geodesic,
their signs dictate whether each respective coordinate
increases or decreases as proper time progresses forward.
In particular, above the outer horizon, the radial Hamilton-Jacobi parameter
$P_x$ can either be negative (corresponding to infalling particles)
or positive (corresponding to outfalling particles),
while the time parameter $P_t$ is always negative
(since particles must move forward in time $t$).
%The Hamilton-Jacobi equation~(\ref{HamiltonJacobihor})
%requires that at horizons,
%where the horizon function $\Delta_x$ passes through zero,
%the combination $- P_t^2 + P_x^2$ of Hamilton-Jacobi parameters
%also passes through zero.
%The time and radial Hamilton-Jacobi parameters $P_t$ and $P_x$ must themselves
%vary continuously across horizons.
%As expounded in the paragraph following equation~(\ref{HamiltonJacobihor}),
%horizons demarcate regions where the Hamilton-Jacobi parameters $P_t$ and $P_x$
%take certain signs.
But in the Black Hole region of the Kerr geometry,
between the outer and inner horizons,
the radial Hamilton-Jacobi parameter $P_x$ must always be negative (infalling),
signifying that space is falling inward faster than light,
while the time parameter $P_t$
can be either positive (outgoing, up-rightward in Penrose diagram~\ref{penrosemassinflation}) or negative (ingoing, up-leftward in Penrose diagram~\ref{penrosemassinflation}).
The expression~(\ref{momentumtetrad}) for the 4-momentum of a particle
in the Boyer-Lindquist tetrad frame shows that,
at the inner horizon,
outgoing ($P_t > 0$) and ingoing ($P_t < 0$) geodesics see each other
infinitely blueshifted.
The infinite blueshift comes from the diverging denominator
$1/\sqrt{|\Delta_x|}$
in the time and radial components $p_t$ and $p_x$ of the tetrad-frame momentum.
%, equation~(\ref{momentumtetrad}).

As outgoing and ingoing streams approach each other near the inner horizon,
they see each other more and more blueshifted,
the blueshift diverging as the reciprocal of the horizon function,
$1/|\Delta_x|$.
At the inner horizon, where the horizon function $\Delta_x$ goes to zero,
the blueshift becomes infinite,
whereupon the streams exceed the speed of light through each other,
and fall through separate outgoing and ingoing inner horizons
into two separate regions of spacetime, the outgoing and ingoing wormholes
in the Penrose diagram~\ref{penrosemassinflation}.
Time $t$ progresses in opposite directions in the two wormholes,
backward in the outgoing wormhole ($P_t > 0$),
forward in the ingoing wormhole ($P_t < 0$).

As long as the spacetime is a complete vacuum,
as the Kerr solution assumes,
there is no contradiction to outgoing and ingoing frames exceeding
the speed of light through each other.
But as soon as there are even the tiniest amounts of outgoing and ingoing
light or matter,
it is impossible for the outgoing and ingoing streams to exceed
the speed of light through each other.
In any real astronomical black hole,
light and matter accreted from outside the horizon
will fuel both outgoing (prograde) and ingoing (retrograde) streams
at the inner horizon.
Inevitably, the Kerr solution must break down at the inner horizon
of any real black hole.

At the outer horizon, where the horizon function $\Delta_x$ also
reaches zero, one may wonder why a similar breakdown does not occur
as it does at the inner horizon thanks to the diverging denominator
$1/\sqrt{|\Delta_x|}$ in expression~(\ref{momentumtetrad}).
%However, there is no such breakdown at the outer horizon.
In the outside Universe,
time $t$ progresses forward,
equivalent to a negative Hamilton-Jacobi time parameter, $P_t < 0$.
Continuity across the outer horizon requires $P_t$ to remain negative there,
meaning that all geodesics are necessarily ingoing as they cross the horizon.
No outgoing geodesics cross the outer horizon.
Another way of thinking about the distinction is that light\-rays near
the event horizon tend to move away from it,
while light rays near the inner horizon are driven closer to it.
The event horizon is a repeller of light rays, the inner horizon an attractor.

The back-reaction on the Kerr geometry caused by counter-streaming
of outgoing and ingoing streams near the inner horizon
was eventually calculated in a seminal paper
by Poisson \& Israel in 1990 \cite{Poisson:1990eh}.
Their first paper concerned spherical charged (Reissner-Nordstr\"om)
black holes,
but they soon followed with a paper on rotating black holes
\cite{Barrabes:1990}.
They showed that the back-reaction was such as to cause an exponential
growth in the interior mass, a phenomenon they dubbed ``mass inflation.''
The proper energy density in the center-of-mass frame of the
counter-streaming streams also exponentiates,
as does the Weyl curvature (tidal force).

As reviewed by \cite{McMaken:2021},
there has been some controversy as to the final outcome of mass inflation.
The original scenario contemplated by Poisson \& Israel was
that the crossflow of outgoing and ingoing streams would be driven
by a ``Price tail'' \cite{Price:1972} of gravitational radiation
generated when the black hole first collapsed.
As confirmed by many subsequent researchers
(see e.g.\ \cite{Dafermos:2025} and references therein),
the outcome in this scenario is a
``weak, null singularity.''
The singularity is weak in the sense that,
although the tidal force diverges to infinity,
it diverges in such a short proper time that objects in the spacetime
are scarcely distorted.

However, as pointed out by \cite{Hamilton:2008zz},
the Price tail of gravitational waves decays too rapidly
to be the primary source of back-reaction.
In any real black hole, accretion of matter or radiation
(cosmic microwave background photons if nothing else)
soon (in a matter of seconds for a stellar-mass black hole)
dominates the outgoing and ingoing streams impinging on the inner horizon.
%and produces energies above the Planck scale.
As long as there is ongoing accretion,
the outcome is collapse to a spacelike singularity.
Numerical calculations of an accreting, rotating black hole
\cite{Hamilton:2017qls}
indicate that the collapse is of the chaotic variety discovered
by Belinskii, Khalatnikov \& Lifshitz in 1970
\cite{Belinskii:1970,Belinskii:1982}.

Recent calculations
\cite{Zilberman:2022a,Zilberman:2022b}
using the apparatus of quantum field theory in curved spacetime
show that even in a pure vacuum in the absence of outside accretion,
quantum effects arising from a Kerr black hole's own gravity
destabilize the inner horizons.
The renormalized expectation value of the quantum energy-momentum tensor
remains finite and well-behaved at the outer horizon,
but it diverges at both outgoing and ingoing inner horizons.
Calculations of the Hawking radiation seen by an infaller
\cite{McMaken:2024}
point to the same conclusion.
As the infaller approaches the black hole and passes the outer horizon,
they continue to experience a tiny amount of Hawking radiation
(we continue to use that term here,
even though it most commonly refers to the asymptotically distant
radiation predicted by \cite{Hawking:1975b}).
But once the infaller reaches either inner horizon,
they see a diverging amount of nonthermal radiation
\cite{McMaken:2024}.
An outgoing infaller sees a diverging ingoing Hawking flux,
while an ingoing infaller sees a diverging outgoing flux.
The results continue to hold in the limit of a spherical (Schwarzschild)
black hole, whose inner horizon is at zero radius;
although curiously enough, the calculations are most challenging
in the limit of small spin,
the series expansions requiring more terms to achieve convergence.

The bottom line of these deliberations is that,
although the Kerr line-element provides an excellent approximation
to the geometry of a real rotating black hole
down to near its inner horizon,
that approximation breaks down dramatically just above the inner horizon.
Counter-propagating outgoing and ingoing streams,
generated either by accretion from outside,
or by the black hole's own quantum self-irradiation,
become hyper-relativistic as they approach the inner horizon,
driving the center-of-mass density and curvature to exponentially huge values.
Classically,
the likely outcome \cite{Hamilton:2017qls,McMaken:2021}
is Belinskii-Khalatnikov-Lifshitz collapse
to a spacelike singular surface at the inner horizon.
A more complete understanding awaits a consensus theory of quantum gravity.

\subsection{Geodesics}
\label{kerrgeodesics-sec}

In a rotating black hole, as in a spherical black hole,
the boundary between regions visible and invisible to an infaller
between the horizons is set by photons that have the largest angular motion.
The Carter constant $\KCarter$,
which for a spherical black hole equals the total angular momentum squared,
is infinite for photons with the largest angular motion,
\begin{equation}
\label{Kborder}
  \KCarter
  =
  \infty
  \ .
\end{equation}
The ratio $J$ of azimuthal to total angular momentum is
\begin{equation}
\label{Jborder}
  J
  \equiv
  \frac{L}{\sqrt{\KCarter}}
  =
  \sin\theta_J
  =
  \pm \sqrt{1 - y_J^2}
  \ ,
\end{equation}
where $y_J \equiv - \cos\theta_J$
is the largest latitude reached on the geodesic
($J = 0$ and $y_J = \pm 1$ for geodesics that reach the poles,
$J = \pm 1$ and $y_J = 0$ on the equator).
%NO
%Positive $J$ is prograde,
%negative $J$ retrograde.
In the limit of infinite $\KCarter$,
the Hamilton-Jacobi solution~(\ref{dxdyP}) simplifies to
\begin{align}
  -
  \int_{x_0} \frac{\dd x}{P_x}
  &=
  \frac{1}{\sqrt{\KCarter}}
  \int_{r_0} \frac{\dd r}{\sqrt{( r - e_- ) ( e_+ - r )}}
\nonumber
\\
  &=
  \frac{2}{\sqrt{\KCarter}}
  \left[
  \atan \sqrt{\frac{r - e_-}{e_+ - r}}
  \right]_{r_0}^r
\nonumber
\\
  =
  \int_{y_0} \frac{\dd y}{P_y}
  &=
  \frac{1}{\sqrt{\KCarter}}
  \int_{y_0} \frac{\dd y}{\sqrt{( y + y_J ) ( y_J - y)}}
\nonumber
\\
  &=
  \frac{2}{\sqrt{\KCarter}}
  \left[
  \atan \sqrt{\frac{y + y_J}{y_J - y}}
  \right]_{y_0}^y
  \ ,
\end{align}
where $e_\pm$ are the radii of the ergospheres at latitude $\theta_J$,
\begin{equation}
  e_\pm
  =
  M \pm \sqrt{M^2 - Q^2 - a^2 \cos^2\!\theta_J}
  \ .
\end{equation}
Define the angle $\psi$ by
\begin{equation}
  \psi
  \equiv
  -
  \sqrt{\KCarter}
  \int_{x_0} \frac{\dd x}{P_x}
  =
  \sqrt{\KCarter}
  \int_{y_0} \frac{\dd y}{P_y}
  \ ,
\end{equation}
which starts at $\psi = 0$ at the position $x_0$, $y_0$.
A null trajectory passing through an emitter or observer
at radius $r_0 \equiv a \cot(a x_0)$
and latitude $y_0 \equiv - \cos\theta_0$
is given parametrically in terms of the angle $\psi$ by
\begin{subequations}
\begin{align}
\label{ryvis}
  r
  &=
  M - \tfrac{1}{2} ( e_+ - e_- ) \cos( \psi + \psi_r )
\nonumber
\\
  &=
  e_-
  +
  ( e_+ - e_- )
  \sin^2\left(\frac{\psi + \psi_r}{2}\right)
  \ ,
\\
  y
  &=
  -
  y_J
  \cos( \psi + \psi_y )
  \ ,
\end{align}
\end{subequations}
where
\begin{subequations}
\begin{align}
  \psi_r
  &\equiv
  %2 \asin\sqrt{\frac{r_0 - e_-}{e_+ - e_-}}
  %=
  \acos\left( - \frac{2(r_0 - M)}{e_+ - e_-} \right)
  \ ,
\\
  \psi_y
  &\equiv
  \acos\left(- \frac{y_0}{y_J}\right)
  \ .
\end{align}
\end{subequations}
The radius $r$ and polar height $y$ vary periodically with the angle $\psi$,
with period $2\pi$.

\kerrvisiblefig

Figure~\ref{kerrvisible} illustrates trajectories in radius $r$
and latitude $\theta$ (azimuth $\phi$ being projected to 0)
for an infaller who reaches the inner horizon at latitude
$\theta_0 = \tfrac{1}{4}\pi$,
equivalent to inclination angle
$\tfrac{1}{2}\pi - \theta_0 = \tfrac{1}{4}\pi$ above the equator.

Compared to the case of a nonrotating (Schwarzschild) black hole
shown in Fig.~\ref{schwJoovis},
visibility inside a rotating black hole is complicated
by the fact that the rotation drags frames around it.
The Boyer-Lindquist time and azimuthal coordinates $t$ and $\phi$
along a geodesic, equations~(\ref{dtdphiP}),
are not the best choice at or inside the horizon,
because they diverge at the horizon thanks to the $1/\Delta_x$ factor.
A better choice of time and azimuthal coordinates
are principal null coordinates $t_\pn$ and $\phi_\pn$ that corotate with the black hole
along the ingoing or outgoing principal null congruences,
\begin{equation}
\label{tphipn}
  t_\pn
  \equiv
  t \pm \int \frac{\dd x}{\Delta_x}
  \ , \quad
  \phi_\pn
  =
  \phi \pm \int \frac{\omega_x \, \dd x}{\Delta_x}
  \ ,
\end{equation}
where the lower ($-$) sign is ingoing, the upper ($+$) sign outgoing.
The corotating coordinates $t_\pn$ and $\phi_\pn$,
also known as Kerr-Eddington-Finkelstein coordinates
(often symbolized $u$/$v$ and $\tilde{\phi}$),
are constant along the ingoing ($-$) or outgoing ($+$) principal null geodesics.
The Boyer-Lindquist time coordinate $t$, which is spacelike between horizons,
decreases along ingoing geodesics inside the outer horizon,
per the Penrose diagram~\ref{penrosemassinflation}.
The corotating time coordinate $t_\pn$
similarly decreases along ingoing geodesics;
indeed it decreases not just between the horizons,
but between the outer and inner ergospheres.
Along outgoing geodesics,
the ingoing corotating coordinate $t_\pn$ decreases between horizons,
diverges at horizons, and increases outside horizons.
The corotating azimuthal coordinate $\phi_\pn$
decreases or increases in tandem with $t_\pn$.

Between outer and inner horizons,
the Hamilton-Jacobi parameter $P_t$ can be either positive (outgoing)
or negative (ingoing).
The $t$ component $p_t = P_t / (\rho \sqrt{|\Delta_x|})$
of the tetrad-frame momentum, equation~(\ref{momentumtetrad}),
diverges at horizons, where $\Delta_x = 0$.
At horizons, outer or inner,
ingoing and outgoing frames stream through each other at the speed of light;
ingoing and outgoing frames see each other infinitely blueshifted.
The divergence occurs only at horizons;
ingoing and outgoing frames that lie strictly between horizons
stream through each other at less than the speed of light.
As described in \S\ref{massinflation-sec},
%In a real black hole formed from gravitational collapse,
%infallers do not encounter an actual divergence at the outer horizon.
all objects that fall through the outer horizon are necessarily ingoing,
so no divergence occurs there.
%Table~\ref{Ptab}.
However, ingoing infallers with sufficiently positive angular momentum $L$ (prograde)
can turn around between the horizons and become outgoing at the inner horizon;
and ingoing objects can emit outgoing light from below the outer horizon.
Light, whether ingoing or outgoing, tends to move away from the
outer horizon, so light does not collect there.
By contrast, ingoing and outgoing light always converges toward the inner horizon.
Even the tiniest streams of ingoing and outgoing light at the inner
horizon ignite the mass inflation instability there.

For definiteness, consider an infaller who is ingoing when they reach
the inner horizon.
The ingoing infaller sees ingoing light to have finite blue/redshift.
Ingoing light with infinite $\KCarter$ necessarily has negative
azimuthal angular momentum $J$
(and outgoing light has positive $J$).
Ingoing corotating coordinates $t_\pn$ and $\phi_\pn$ along a geodesic are,
from equations~(\ref{dtdphiP}),
\begin{subequations}
\label{tphipnm}
\begin{align}
\label{tpnm}
  t_\pn
  &=
  \int_{x_0}
  \left( \frac{P_t}{P_x} - 1 \right) \frac{\dd x}{\Delta_x}
  +
  \int_{y_0}
  \frac{P_\phi}{P_y} \frac{\omega_y \, \dd y}{\Delta_y}
  \ ,
\\
\label{phinm}
  \phi_\pn
  &=
  \int_{x_0}
  \left( \frac{P_t}{P_x} - 1 \right) \frac{\omega_x \, \dd x}{\Delta_x}
  +
  \int_{y_0}
  \frac{P_\phi}{P_y} \frac{\dd y}{\Delta_y}
  \ .
\end{align}
\end{subequations}
If outgoing instead of ingoing frames were considered,
then outgoing coordinates $t_\pn$ and $\phi_\pn$ should be used,
with a $+$ instead of a $-$ sign in equations~(\ref{tphipnm}).
%and the signs of $t_\pn$ and $\phi_\pn$
%in the expressions~(\ref{tphivis}) below
%should be flipped.

Abbreviate
\begin{subequations}
\begin{align}
\label{repm}
  r_\diamond
  \equiv
  \tfrac{1}{2} ( r_+ - r_- )
  &=
  \sqrt{M^2 - Q^2 - a^2}
  \ ,
\\
  e_\diamond
  \equiv
  \tfrac{1}{2} ( e_+ - e_- )
  &=
  \sqrt{M^2 - Q^2 - a^2 y_J^2}
  \ ,
\end{align}
\end{subequations}
and
\begin{subequations}
\begin{align}
\label{csry}
  c_r
  \equiv
  \cos ( \psi + \psi_r )
  \ ,& \quad
  s_r
  \equiv
  \sin ( \psi + \psi_r )
  \ ,
\\
  c_y
  \equiv
  \cos ( \psi + \psi_y )
  \ ,& \quad
  s_y
  \equiv
  \sin ( \psi + \psi_y )
  \ .
\end{align}
\end{subequations}
The polar contribution to $t_\pn$, equation~(\ref{tpnm}),
can be incorporated into the radial contribution by using
$- \dd x / P_x = \dd y / P_y$ to recast it as
\begin{equation}
  \int_{y_0}
  \frac{P_\phi \omega_y}{\Delta_y} \, \frac{\dd y}{P_y}
  =
  -
  \int_{x_0}
  \frac{P_\phi \omega_y}{\Delta_y} \frac{\dd x}{P_x}
  =
  -
  \int_{x_0}
  a J \frac{\dd x}{P_x/\sqrt{\KCarter}}
  \ ,
\end{equation}
valid in the limit of infinite $\KCarter$,
when $P_\phi = L = \sqrt{\KCarter} J$.
The time and azimuthal angles $t_\pn$ and $\phi_\pn$
along null geodesics with $\KCarter = \infty$
integrate to
\begin{subequations}
\label{tphivis}
\begin{align}
\label{tvis}
  t_\pn
  &=
  -
  \left(
  \pm \xi - \xi_0
  +
  2 \xi_\diamond i_r
  \right)
  \ ,
\\
\label{phivis}
  \phi_\pn
  &=
  -
  \left(
  \pm \alpha - \alpha_0
  +
  2 \alpha_\diamond i_r
  +
  \beta - \beta_0
  \right)
  \ ,
\end{align}
\end{subequations}
in which overall minus ($-$) signs have been introduced because,
as described following equations~(\ref{tphipn}),
$t_\pn$ and $\phi_\pn$
decrease along ingoing geodesics between the outer and inner ergospheres;
for outgoing geodesics with respect to outgoing $t_\pn$ and $\phi_\pn$,
the overall signs in equations~(\ref{tphivis}) would be plus ($+$).
The $r$-dependent functions $\xi$, $\alpha$ in equations~(\ref{tphivis})
vary periodically with $\psi + \psi_r$, with period $2\pi$.
The $\pm$ signs in $\pm \xi$ and $\pm \alpha$ in equations~(\ref{tphivis})
are the sign of
$\sin(\psi + \psi_r)$,
and the integer $i_r \equiv [ ( \psi + \psi_r ) / \pi ]$
is the $\pi$ phase of $\psi + \psi_r$,
which increments by 1 at the same time as the $\pm$ sign of
$\pm \xi$ or $\pm \alpha$ flips sign,
ensuring that $t_\pn$ and $\phi_\pn$ vary smoothly as $\xi$ and $\alpha$
pass through their radial extrema.
The functions $\xi$, $\alpha$, and $\beta$ in equations~(\ref{tphivis}) are
\begin{subequations}
\label{xivis}
\begin{align}
  \xi
  &\equiv
  \sign(c_r)
  \biggl(
  \frac{2 M^2 - Q^2}{2 r_\diamond}
  \ln
  \left(
  \frac{e_\diamond + a |J| |s_r| - r_\diamond |c_r|}{e_\diamond + a |J| |s_r| + r_\diamond |c_r|}
  \right)
\\
  &\quad
  +
  e_\diamond |c_r|
  \biggr)
  -
  2 M \ln \left( 1 + \frac{e_\diamond |s_r|}{a |J|} \right)
  \ ,
\\
  \xi_\diamond
  &\equiv
  \frac{2 M^2 - Q^2}{r_\diamond}
  \ln \left( \frac{e_\diamond + r_\diamond}{a |J|} \right)
  -
  e_\diamond
  \ ,
\\
  \alpha
  &\equiv
  \sign( c_r )
  \frac{a}{r_\diamond}
  \ln
  \left(
  \frac{r_\diamond |s_r| + a |J| |c_r|}
  {r_\diamond + e_\diamond |c_r|}
  \right)
  \ ,
\\
  \alpha_\diamond
  &\equiv
  \frac{a}{r_\diamond}
  \ln \left( \frac{r_\diamond + e_\diamond}{a |J|} \right)
  \ ,
\\
  \beta
  &\equiv
  \acot \left(
  |J| c_y , s_y
  \right)
  \ .
\end{align}
\end{subequations}
The $r$-dependent functions $\xi$ and $\alpha$ have minima respectively
$\xi = - \xi_\diamond$
and
$\alpha = - \alpha_\diamond$
at $\psi + \psi_r = 0$ where $r = e_-$,
pass through $0$
at $\psi + \psi_r = \tfrac{1}{2}\pi$ where $r = M$,
and reach maxima
$\xi = \xi_\diamond$
and
$\alpha = \alpha_\diamond$
at $\psi + \psi_r = \pi$ where $r = e_+$.
At outer or inner horizons, $s_r = \pm a |J| / e_\diamond$
and $c_r = \pm \sqrt{1 - s_r^2}$,
with signs set by the phase of $\psi + \psi_r$.
The sequence as $\psi + \psi_r$ increases from 0 to $2\pi$ is,
inner horizon at $\psi+\psi_r \in [0,\tfrac{1}{2}\pi]$,
outer horizon at $\psi+\psi_r \in [\tfrac{1}{2}\pi,\pi]$,
outer horizon at $\psi+\psi_r \in [\pi,\tfrac{3}{2}\pi]$,
inner horizon at $\psi+\psi_r \in [\tfrac{3}{2}\pi,2\pi]$.
The values of $\xi$ and $\alpha$ at outer and inner horizons are
\begin{subequations}
\label{xivishor}
\begin{align}
  \xi_\pm
  &\equiv
  \pm
  \left(
  \frac{2 M^2 - Q^2}{r_\diamond}
  \ln
  \left(
  \frac{e_\diamond}{a |J|}
  \right)
  -
  r_\diamond
  \right)
  -
  2 M \ln 2
  \ ,
\\
  \alpha_\pm
  &\equiv
  \pm
  \frac{a}{r_\diamond}
  \ln
  \left(
  \frac{e_\diamond}{a |J|}
  \right)
  \ ,
\end{align}
\end{subequations}
where the upper and lower signs are at the outer and inner horizons,
respectively.
Note that the parts of the sequence inside the inner horizon
do not occur in a real rotating black hole,
because the mass inflation instability invalidates the Kerr solution there.

An observer at $r_0$ watching an emitter at $r$
sees the emitter as they used to be a time $t_\pn$ ago.
During that time the corotating coordinate $\phi_\pn$ at $r$
has changed by $\omega_x t_\pn$,
so the observer sees the emitter at azimuthal coordinate
\begin{equation}
  \phi_\pn - \omega_x t_\pn
  \ .
\end{equation}
In particular, an ingoing observer at the inner horizon $r_-$
watching an ingoing light ray with infinite $\KCarter$ falling through
the outer horizon $r_+$ sees the emitter at azimuthal coordinate
\begin{align}
\label{dphipnhor}
  &\phi_\pn - \omega_x t_\pn
\nonumber
\\
  &=
  -
  \left(
  ( \alpha_+ - \alpha_- )
  +
  ( \beta_+ - \beta_- )
  - \frac{a}{r_+^2 + a^2} ( \xi_+ - \xi_- )
  \right)
\nonumber
\\
  &= 
  \frac{4 a}{r_+^2 + a^2}
  \left(
  M
  \ln
  \left(
  \frac{e_\diamond}{a |J|}
  \right)
  -
  r_\diamond
  \right)
  -
  ( \beta_+ - \beta_- )
\end{align}
relative to their own azimuthal coordinate $\phi_\pn$.
The relative azimuthal coordinate~(\ref{dphipnhor})
is not coordinate gauge-invariant,
but the difference in the perceived azimuthal coordinates of different
emitters at the same outer horizon radius $r_+$
is coordinate gauge-invariant.
A takeaway from equation~(\ref{dphipnhor}) is that
the perceived azimuthal coordinate
$\phi_\pn - \omega_x t_\pn$
diverges logarithmically as $|J| \rightarrow 0$,
indicating that an observer at the inner horizon
sees geodesics with tiny $J$ (and infinite $\KCarter$)
circulate the black hole many times,
dragged around by the black hole's rotation.
Numerical experiment confirms that the boundary of the (in)visible
regime is determined by geodesics with the smallest azimuthal
angular momentum, $J \sim 0$.
The geodesics hug the $J = 0$ surface in radius $r$ and latitude $y$,
spiraling around in azimuth $\phi_\pn$,
all the way from outer to inner horizon.
The resulting (in)visible region is illustrated in Figure~\ref{kerrvisible}.

\section{Implications for quantum gravity}
\label{qg-sec}

The conceptual adjustment
that the singularity of a black hole is a surface, not a point,
has repercussions for quantum gravity.
%The singular surface resides deep inside the black hole.
%We claim that this challenges some prevailing ideas about the quantum nature
%of black holes.

Without committing to any particular theory of quantum gravity,
this section~\ref{qg-sec} advances four claims about its desired features:
%The purpose of this section is not to put forward or advocate
%any particular theory of quantum gravity,
%but rather to point out the probable features
%that a quantum theory of gravity should have,
%if indeed the singularity is a surface.
%It should be emphasized that, absent a final theory of quantum gravity,
%one cannot be certain that the arguments that follow are true.
%However, the story does seem to hold together nicely.
%
%The key idea this section seeks to defend is modest and theory-agnostic,
%yet precise: there exists a quantum system deep within a black hole's
%interior whose coarse-grained geometric description collapses
%to a spacelike 2D surface in the classical limit.
%If this is correct, then,
%as will be argued in \S\ref{unitary-sec}:
\begin{enumerate}
\item
The quantum states of a black hole reside at its singular surface,
not at its horizon.
\item
Quantum mechanics requires that the singular surface evolve unitarily.
\item
The singular surface coevolves unitarily with its own Hawking radiation.
\item
The singular surface probably evolves rapidly toward thermodynamic equilibrium.
\end{enumerate}

We emphasize that we are not advocating that the singular surface should,
in a full theory of quantum gravity,
be interpreted as literally a 2-dimensional geometric surface.
In string theory, for example, there is a fundamental length scale,
and there are T-dualities that exchange small and large scales.
In loop quantum gravity, spacetime resolves into a discrete spin network.
%Rather, we envisage that whatever quantum object replaces the classical
%singular boundary should admit an effective coarse-grained description
%as a 2 dimensional surface,
%as Carlip \cite{Carlip:1999} has proposed.
Whatever thing the singular surface resolves into,
%In a complete theory of quantum gravity,
%the singular surface should somehow become a real quantum thing,
%(an oozlum, for lack of a better term),
it will possess a Hilbert space of states,
along with equations that describe its unitary evolution.
The present paper is agnostic as to what those states and equations might be.

%The discussion of these implications is followed in \S\ref{string-sec} by brief comments on how they might fit within the framework of two common approaches to theories of quantum gravity, holography and string theory.

%We emphasize that the views expressed in this section~\ref{qg-sec} are our own,
%and do not represent a consensus of the community.
%Our views are colored by our own and others' calculations
%\cite{Zilberman:2022a,Zilberman:2022b,McMaken:2024}
%of diverging quantum energy-momentum and Hawking radiation
%at the inner horizons of black holes
%(including spherical black holes, where the inner horizon is at zero radius,
%$r = 0$).
%These calculations, carried out
%using the established apparatus of quantum field theory in curved spacetime,
%should be robust
%regardless of whatever the ultimate theory of quantum gravity might be.

\subsection{The quantum states of a black hole reside at its singular surface}
\label{quantumstates-sec}

The states of a black hole are commonly conceived to reside,
or at least to be encoded somehow, on its horizon.
The idea traces to Hawking \cite{Hawking:1975b},
who first showed that a black hole behaves like a thermodynamic object
with entropy equal to $1/(4G)$ times the area of its horizon.
%in Planck units $c = \hbar = G = 1$.
In statistical mechanics,
the entropy counts the logarithm of the number of quantum states
occupied by the system.
The notion that a black hole's entropy really can be treated in this
statistical mechanical way gains support from the fact
%these are actual quantum states gains support from the fact
that leading theories of quantum gravity,
such as string theory, e.g.\ \cite{Strominger:1996b},
and loop quantum gravity, e.g.\ \cite{Ashtekar:1998},
are able to reproduce the correct formula for entropy.

%For a black hole, the required number of states grows exponentially
%with the mass squared of the black hole, $N \sim \ee^{M^2}$.
%In string theory in particular,
%the number of massive states of excited strings grows exponentially.
%Susskind et al.\ \cite{Susskind:1993if} argued that
%for a system of specified mass, the partition function is dominated
%by the massive excitations of a single long string,
%to be thought of as somehow wrapped around the horizon.

However, a prominent takeaway from the present paper is that someone
freely falling through the event horizon does not encounter anything untoward there.
An observer looking at an astronomical black hole from the outside
is not looking at its event horizon.
Rather, they are looking at the exponentially dimming, redshifting surface
of the star that collapsed long ago,
which \cite{Hawking:1973} called the past horizon,
and \cite{Hamilton:2010my} dubbed the illusory horizon.
When an observer falls through the event horizon,
they do not catch up with the collapsed star,
but rather continue to see it ahead of them,
still redshifting (or blueshifting) away,
as depicted in Figure~\ref{schwviz}.

%As seen in the visualization in Figure~\ref{schwviz},
%the appearance of the past horizon is continuous throughout the observer's
%inward voyage, from outside the event horizon, through the event horizon,
%down to the point where the observer is about to hit the singularity,
%when they have the impression of finally encountering the singular surface.

Since the observer near the singularity can perceive only the part
of spacetime lying within their past lightcone,
their view asymptotically kisses the singular surface
at just a single point in the spatial $\{ t, \theta, \phi \}$ directions.
The rest of the singular surface remains hidden from view,
behind the past horizon.

In quantum field theory,
the statistical state of a system in which an observer
has incomplete information is described by a density matrix
acting on the Hilbert space of states.
Associated with the density matrix is an entropy,
obtained by tracing over the observer's ignorance.
This suggests that the entropy of the black hole is a trace over
the unseen states at the singular surface,
and this remains true not only for observers outside the black hole
but also for those deep inside,
down to where they are about to hit the singular surface.

If the states of a black hole reside at its singular surface,
what does it mean to say that the states are encoded on its horizon?
Carlip \cite{Carlip:1999}
has argued that
the entropy of any local Killing horizon whose transverse oscillations satisfy
a conformal symmetry in the orthogonal $t$--$r$ direction
reproduces the Bekenstein-Hawking entropy formula.
He argues that
this could explain the simplicity and universality of the entropy formula.
He does not require that microscopic states be physically localized to the horizon;
a (near-)horizon symmetry,
without reference to its source,
is sufficient to reproduce the entropy.
%Carlip treats the surface as being the horizon.
%But it is worth emphasizing that Carlip's mathematical derivation
%does not require the surface actually be the horizon:
%rather, it requires that the Hilbert space of the black hole
%be approximated by the excitations of a 2~dimensional surface
%having conformal symmetry.

Modern approaches to holography usually discard the possibility that
information on the horizon is encoded solely by interior degrees of freedom
because of the mismatch between the entropy's area scaling and
the bulk interior's volume scaling,
which can grow arbitrarily large without substantially changing the horizon area.
Instead, degrees of freedom are usually treated as being encoded nonlocally
(via, e.g., entanglement wedge reconstruction),
rendering the question of where states reside meaningless in a gauge-invariant sense.
The presence of nonlocal, correlational degrees of freedom is explored further
in \S\ref{hawking-sec}.
But this paper suggests that the more important source of quantum states
lies not throughout the volume of the bulk but rather confined to the
lower-dimensional structure of the singularity.
This claim is supported by the recent calculations of
\cite{Zilberman:2022a,Zilberman:2022b,McMaken:2024},
which show that the quantum energy-momentum and Hawking radiation seen by an infaller
remain finite and tiny at the outer horizon and throughout the bulk,
but diverge at the inner horizon,
where, as argued in \S\ref{massinflation-sec},
the Poisson-Israel inflationary instability likely precipitates collapse
to a singular surface.

\subsection{Unitary evolution of the singular surface}
\label{unitary-sec}

In general relativity,
the singularity of the Schwarzschild geometry is a boundary beyond
which the geometry cannot be continued.
The singularity is the 3-dimensional boundary of 4-dimensional spacetime.
The dimension that comes to an end at the singularity is the time
(the timelike radial) dimension, so the singular boundary is spacelike.

This cannot be true in quantum gravity.
%Time cannot simply ``come to an end.''
A fundamental tenet of quantum mechanics is that evolution must be unitary.
Quantum mechanics deals with probabilities of quantum states at one time
evolving into quantum states at another time.
Unitary evolution means that amplitudes of states at one time are
related by a unitary transformation to amplitudes at another time,
which ensures that the sum of probabilities (squared amplitudes)
is always one.
Presumably quantum gravity replaces the
%``exitium in nihilum'' (opposite of ``creatio ex nihilo'')
{\em exitium in nihilum\/} (``exit into nothing'')
predicted by general relativity by ``something,''
and that something should evolve unitarily.

Suppose that the black hole has collapsed and settled down.
Unitarity requires that infallers who subsequently fall in
and encounter the singular surface
at different times $t$ and angular locations $\theta,\phi$
should count the same set of quantum states.
If unitarity holds,
the singular surface should be not a 3-dimensional classical boundary,
but rather an effective 2-dimensional surface that evolves unitarily.

Although classically the time coordinate $t$ is spacelike at the
singular boundary,
it does provide a natural time ordering of events at the singularity,
insofar as objects that fall to the singularity at later times $t$
probe the singularity at later times than objects that fell in earlier.

We do not offer a proof of unitarity, which is notoriously tricky.
Rather, we assert that unitarity is a necessary desideratum
for a successful quantum description of a black hole's singular surface.

\subsection{Hawking radiation as a mediator of unitary evolution}
\label{hawking-sec}

Recent calculations using the apparatus of quantum field theory in
curved spacetime
\cite{Zilberman:2022a,Zilberman:2022b,McMaken:2024}
reveal diverging quantum energy-momentum and Hawking radiation
at the inner horizons of black holes
(including spherical black holes, where the inner horizon is at zero radius,
$r = 0$).
%Asymptotically close to the singularity of a spherical black hole
%or the inner horizon of a rotating black hole,
%the quantum energy-momentum and Hawking radiation fluxes in the Unruh state diverge
%\cite{Zilberman:2022a,Zilberman:2022b,McMaken:2024}.
This fact suggests that the singular surface should be thought of
not as having infinitesimal thickness, but rather
as being accompanied by a roiling atmosphere of Hawking radiation.
%The singular surface and its Hawking atmosphere must evolve unitarily together.
The situation is reminiscent of that in quantum field theory,
where a point particle such as an electron or photon
should be thought of not as an isolated point,
but rather as accompanied by a roiling cloud of virtual particles
polarized out of the vacuum by the presence of the particle.

Hawking radiation is a natural mediator of unitary evolution.
%of the singular surface.
A Hawking pair, drawing its energy from the black hole's curvature,
is created in a maximally entangled state.
One of the pair
%, with negative energy measured relative to infinity,
falls to the singular surface,
while its partner
%, with positive energy,
either escapes the black hole,
or else also falls to the singular surface.
The singularity ``observes'' the first of the pair,
but that observation does not destroy the entanglement,
but rather shares the entanglement with other states of the singularity.
If the partner escapes to the outside where it is observed,
then likewise that observation does not destroy the entanglement,
but rather shares the entanglement with other states of the outside system,
so that the singularity and outside world remain entangled with each other.
If, instead of escaping, the partner falls to another place on the singular
surface, then the partner shares its entanglement with that region
of the singular surface, thereby entangling different spacelike-separated
parts of the singular surface.
When another Hawking pair is created,
it repeats the process of further entangling.
In this fashion the Hawking radiation that escapes the black hole
becomes thoroughly entangled not only with the interior of the black hole,
but also with other Hawking radiation that escapes.
%In this picture, there is no information problem
%demanding an effective description with horizon-scale quantum states.
%The singular surface evolves locally,
%
%The entire system evolves unitarily,
%with time effectively replaced by correlations between quantum degrees of freedom.
%Hawking radiation provides the means to illuminate the
%invisible regions calculated in sections~\ref{schw-sec} and \ref{kerr-sec},
%distributing entanglement between spacelike-separated points
%within a strongly interacting quantum system.
%%In this picture, there is no information paradox,
%%because any observation of either one of a Hawking pair,
%%whether inside or outside the black hole,
%%serves to distribute entanglement between spacelike-separated points,
%%but cannot carry information,
%%because the observer can only observe Hawking radiation passively:
%%they cannot influence which of the possible quantum outcomes
%%they will observe.

\subsection{Evolution to thermodynamic equilibrium}

If the singular surface coevolves unitarily with its hot atmosphere of trapped
Hawking radiation,
then any disturbance from the outside should quickly dissipate,
leaving the surface and its Hawking radiation in a condition
of mutual thermodynamic equilibrium.
%The laws of black hole thermodynamics pioneered by
%Bekenstein \cite{Bekenstein:1973ur} and Hawking \cite{Hawking:1975b}
%suggest that black holes are thermodynamic objects,
%appearing from the outside to emit blackbody radiation.
%%as though they were objects in thermodynamic equilibrium.
%However, from the perspective of an infaller,
%the black hole will also appear from the inside to emit radiation,
%suggesting that the fundamental states of the black hole
%reside at, or near to, its singular surface, not at the horizon.

In the AdS-CFT correspondence \cite{Maldacena:1997re},
the reflecting AdS boundary is crucial to reflect a black hole's
Hawking radiation back into the black hole,
allowing the Hawking radiation to come into thermodynamic equilibrium,
and ensuring unitarity.

In the present picture there is no need for a reflecting AdS boundary
to confine the Hawking radiation.
The great bulk of the Hawking radiation is generated inside the event horizon,
and the trapped Hawking radiation is forced back to the singular surface.
The Hawking radiation that dribbles out to the world beyond the event horizon
is but a stifled remnant of the black hole's fiery core.

\section{Conclusion}

%The question of what really happens inside black holes is of
%perennial interest to students and the general public.
%The astronomical evidence that black holes exist,
%and that they conform to the predictions of general relativity,
%grows stronger year by year,
%including the detection of gravitational waves
%from a pair of merging black holes
%\cite{Abbott:2016blz},
%the size and appearance of the M87 black hole
%measured with the Event Horizon Telescope
%\cite{EHT1:2019},
%and the precession of the orbit of the star S2 around the Milky Way
%black hole Sgr~A$^\ast$ \cite{Gravity:2020}.
%Meanwhile, the singularity theorems pioneered by
%\cite{Penrose:1965}
%and Hawking
%\citep{Hawking:1973}
%are persuasive that singularities are inevitable in general relativity
%(see e.g.\ the review by \citep{Senovilla:1997}),
%signaling a need for new physics.
%
%Although there is a large literature on quantum gravity of all stripes,
%there is surprising little literature on the question of what really happens
%inside real astronomical black holes.
%In the absence of research, misconceptions have taken root,
%of which the most widespread is
%that the singularity at the center of a black hole is a point.

This paper has argued that the widely repeated statement
that the singularity at the center of a black hole is a point is incorrect.
Although points near the singularity of a Schwarzschild black hole
are spatially close, they are causally distant.
As illustrated in Fig.~\ref{schwJoovis},
when two observers at two different angular positions fall to the singularity,
they do not meet each other at the singularity;
rather, they lose causal contact with each other
already well away from the singularity.
The singularity is a 2-dimensional surface, not a point.

The same is true for rotating black holes.
In an astronomically realistic black hole,
the Kerr geometry provides an accurate approximation down to the
region just above the inner horizon, but it fails dramatically at the inner horizon.
Even the tiniest combination of outgoing and ingoing perturbations,
whether instigated by accretion from outside \cite{Hamilton:2017qls},
or by the black hole's own quantum self-irradiation,
\cite{Zilberman:2022a,Zilberman:2022b,McMaken:2024},
drive the Poisson-Israel \cite{Poisson:1990eh} exponential instability,
precipitating collapse to a spacelike singular surface.
The instability and collapse effectively truncate the Kerr geometry
at its inner horizon.
Infallers who reach the inner horizon at two different angular positions are,
as in a Schwarzschild black hole, causally distant,
Fig.~\ref{kerrvisible}.

An observer looking at a black hole from the outside is not looking
at its event horizon;
rather they are looking at the exponentially dimming, redshifting surface
of the star that collapsed long ago,
which \cite{Hawking:1973} called the past horizon,
and \cite{Hamilton:2010my} dubbed the illusory horizon.
When an observer falls through the event horizon,
they do not catch up with the collapsed star,
but rather continue to see it ahead of them,
still redshifting (or blueshifting) away.
Since the infaller can see only within their past lightcone,
which narrows to a cusp at the singular surface,
the singular surface remains hidden from view behind the past/illusory horizon,
all the way down to the point where the infaller hits the singular surface.
At this point the infaller has the impression
that they catch up with the past/illusory horizon,
which, thanks to the diverging tidal force,
gives the appearance of flattening to a flat surface, Figure~\ref{schwviz}.

Section~\ref{qg-sec} discusses the implications for quantum gravity.
It argues four key points:
(1) regardless of whatever the final theory of quantum gravity might be,
the quantum states of a black hole reside at its singular surface;
(2) the singular surface should evolve unitarily;
(3) it should coevolve with its own trapped Hawking radiation;
and (4) it should evolve rapidly to a state of thermodynamic equilibrium.
An infaller who has not (quite) reached the singular surface
never gets to see the singular surface:
it is always hidden behind the past horizon,
even for infallers asymptotically close to it.
For an infaller,
the entropy of the black hole is a trace over the unseen states
of the singular surface,
which remains true not only for observers outside the event horizon,
but also for those deep inside, down to where the observer
is just about to hit the singular surface.

Calculations using the apparatus of quantum field theory in curved spacetime
\cite{Zilberman:2022a,Zilberman:2022b,McMaken:2024}
show that as an infaller finds themself approaching the inner horizon,
the quantum energy-momentum and Hawking radiation they see diverge,
indicating that the singular surface is hot.
The calculations indicate that the great bulk of Hawking radiation
is generated inside the event horizon;
only a tiny fraction leeks out to the universe outside.
The trapped Hawking radiation is forced inward back to the singular surface.
In AdS-CFT, an AdS boundary is needed to reflect a black hole's
Hawking radiation back to its event horizon,
allowing the black hole to come into thermal equilibrium with its own
Hawking radiation, and enabling the closed system to evolve unitarily.
In the present picture there is no need for an AdS boundary,
because the profuse Hawking radiation generated inside the event horizon
is trapped by the black hole's own gravity.
Any disturbance from outside should be able to dissipate rapidly,
leaving the singular surface and its attendant Hawking radiation in
a condition of mutual thermodynamic equilibrium.
We conjecture that, as in AdS-CFT, the system evolves unitarily,
as quantum mechanics demands.
Treating the black hole singularity as a hot, quantum correlated surface
provides a path forward to understanding the true quantum gravitational
nature of the most energetic places in the known Universe.

\section*{Data availability}

There are no publicly available research data or software
supporting this manuscript.
Requests for further information or data should be sent to the authors.

\appendix   % Omit the * if there's more than one appendix.

\section{Geodesics in the Schwarzschild geometry}
\label{schwarzschildgeodesics-sec}

\subsection{Geodesics}
\label{schwgeodesics-sec}

The Schwarzschild line-element for a black hole of mass $M$ is,
in terms of Schwarzschild time $t$ and polar coordinates $r$, $\theta$, $\phi$
(units $c = 1$),
\begin{equation}
\label{schwmetric}
  \dd s^2
  =
  - \,
  \Delta \, \dd t^2
  + {1 \over \Delta} \, \dd r^2
  + r^2 \, 
  ( \dd \theta^2 + \sin^2 \! \theta \, \dd \phi^2 )
  \ ,
\end{equation}
where $\Delta$ is the horizon function
\begin{equation}
  \Delta
  \equiv
  1 - {r_s \over r}
  \ , \quad
  r_s \equiv {2 G M}
  \ .
\end{equation}
With units of the speed of light $c$ restored,
the Schwarzschild radius is $r_s = 2 G M / c^2$.
Without loss of generality, a geodesic can be taken to follow $\phi = 0$.
For a particle of mass $m$,
conservation of energy $E$,
angular momentum $L$,
and rest mass $m$ implies
that the 4-momentum
$p^{\mu} = m \, \dd x^{\mu} / \dd \tau$
along the geodesic satisfies
\begin{equation}
\label{orbiteq}
  p_{t} = -E
  \ , \quad
  p_{\theta} = L
  \ , \quad
  p_{\mu} p^{\mu} = -m^2
  \ .
\end{equation}
Equations~(\ref{orbiteq}) imply
that the radial component $p^{r}$ of the 4-momentum satisfies
\begin{equation}
\label{prorbit}
  p^{r}
  =
  \pm \,
  \left(
  E^2 - U
  \right)^{1/2}
  \ ,
\end{equation}
where $U$ is the effective potential
\begin{equation}
\label{Uorbit}
  U
  \equiv
  \left( m^2 + {L^2 \over r^2} \right)
  \Delta
  \ .
\end{equation}
Equation~(\ref{prorbit}) solves to give the angle $\theta$ as a function
of radius $r$ along the geodesic,
\begin{equation}
\label{thetaorbit}
  \theta
  =
  \int
  {L \, \dd r \over \sqrt{( E^2 - m^2 ) r^4 - L^2 r^2 \Delta}}
  \ .
\end{equation}
For a massless particle, $m = 0$,
geodesics depend only on the ratio
\begin{equation}
  J \equiv \frac{L}{E}
\end{equation}
of the angular momentum to energy.
The maximum possible angular momentum inside the horizon
is $J \rightarrow \infty$,
in which case equation~(\ref{thetaorbit}) simplifies to
\begin{equation}
\label{thetaooorbit}
  \theta
  =
  \int
  {\dd r \over \sqrt{r ( r_s - r )}}
  =
  2 \arcsin \sqrt{r \over r_s}
  \ ,
\end{equation}
which gives the cardioid equation~(\ref{rschwJoo}).

\subsection{Flattening of the singular surface perceived by an infaller}
\label{flattening-sec}

In the locally orthonormal tetrad frame of a timelike observer with the
4-momentum given by equations~(\ref{orbiteq}),
the coordinate-frame 4-momentum $p^{\mu}$ of a photon will transform
into a tetrad-frame 4-momentum $p^{m}=e^{m}{}_{\mu}\ p^{\mu}$,
where $g_{\mu\nu}e_{m}{}^{\mu}e_{n}{}^{\nu}=\eta_{mn}$.
The angle $\chi$ this photon will be observed to make with respect to the
outward radial direction will then satisfy
\begin{equation}
    \tan\chi=\frac{p^{2}}{p^{1}}
    \ .
\end{equation}
For an infalling observer free-falling
with energy $E$ and zero angular momentum, $L = 0$,
watching a light ray with angular momentum per unit energy $J$,
the angle $\chi$ satisfies
\begin{equation}
\label{tetradangle}
  \frac{J}{r}
  =
  \frac{\sin\chi}{(E/m) + \cos\chi \sqrt{(E/m)^2 - \Delta}}
  \ .
\end{equation}
In the limit $r \rightarrow 0$,
where $\Delta \equiv 1 - r_s/r \rightarrow -\infty$,
equation~(\ref{tetradangle}) simplifies to
\begin{equation}
  \tan\chi
  \rightarrow
  \frac{J}{r} \sqrt{\frac{r_s}{r}}
  \rightarrow
  \infty
  \ ,
\end{equation}
that is,
\begin{equation}
\label{tetradangle0}
  \chi \rightarrow \frac{\pi}{2}
  \ ,
\end{equation}
independent of the energy $E$ of the infaller.
Equation~(\ref{tetradangle}) indicates that null rays observed by an infaller
approaching the singularity will become aberrated to the direction perpendicular
to the radial axis, corresponding to a great circle on the observer's sky.
If the observer has a finite angular momentum $L$,
then the observer's transverse motion further aberrates the view in the
transverse direction, but it traces out the same great circle on the sky.
The past horizon thus has the appearance of flattening into an infinite plane,
as illustrated in Figure~\ref{schwviz},
regardless of the state of motion of the infaller.

\subsection{Causal (affine) distance between infallers}
\label{affine-sec}

The proper spatial distance between two infallers who reach the
singularity at the same Schwarzschild time $t$ but
along different radial directions goes to zero at the singularity,
but the causal distance between the two,
the shortest causal path joining them, does not go to zero.
The shortest causal path is illustrated by the red lines
in the right panel of Figure~\ref{schwJoovis},
a pair of null geodesics each with the maximum possible angular momentum,
$J \rightarrow \infty$.

The affine distance $\lambda$ along a geodesic, massive or massless,
is defined by
\begin{equation}
  p^\mu
  =
  m \frac{\dd x^\mu}{\dd \tau}
  =
  \frac{\dd x^\mu}{\dd \lambda}
  \ .
\end{equation}
The affine distance remains finite in the massless limit $m \rightarrow 0$,
and thus provides a measure of causal distance along a null geodesic.
Physically, the affine distance is the proper distance along a light ray
measured by rulers parallel-transported along the light ray
\cite{Hamilton:2010my}.
The affine distance depends on a frame-dependent normalization factor
proportional to the observed frequency $\omega_\obs$ of the light ray.
Affine distances measured in different frames are related by a Lorentz boost.

The affine distance $\lambda$ along a null geodesic is obtained by
integrating
$\dd \theta / \dd \lambda = p^\theta$
or equivalently
$\dd r / \dd \lambda = p^r$.
Normalized to a frame at rest at infinity, the affine distance is
\begin{equation}
\label{lambdaschw}
  \lambda
  =
  {1 \over J}
  \int
  r^2 \, \dd \theta
  =
  \int
  {\dd r \over \sqrt{1 - J^2 \Delta / r^2}}
  \ .
\end{equation}
Normalized to the frame of an observer, the affine distance $\lambda_\obs$ is
\begin{equation}
\label{lambdaobsschw}
  \lambda_\obs
  =
  \omega_\obs \lambda
  \ ,
\end{equation}
where $\omega_\obs$ is the observed frequency (energy)
relative to that at rest at infinity.
For an observer with angular momentum $L_\obs$ at radius $r_\obs$
observing a null geodesic with $J \rightarrow \infty$,
the observed frequency is
\begin{equation}
  \omega_\obs
  =
  \frac{J \sqrt{L_\obs^2 + r_\obs^2}}{r_\obs^2}
  \ .
\end{equation}

The shortest causal path joining infallers reaching the singularity
is realized by a pair of photons emitted in opposite directions
with maximum angular momentum, $J \rightarrow \infty$,
from a point halfway (in angle) between the infallers,
illustrated by the red lines in the right panel of Figure~\ref{schwJoovis}.
The causal path has two symmetrically equal parts,
each following the path of a cardioid, equation~(\ref{rschwJoo}).
If the angular separation between the two infallers near the singularity
is $2\theta$, then the observed affine distance along the shortest causal path
is $2 \lambda_\obs$,
twice the affine distance along each individual null segment,
\begin{align}
\label{lambdashortestcausal}
  2 \lambda_\obs
  &=
  {2 \omega_\obs \over J}
  \!
  \int_0^\theta
  r^2 \, \dd \theta
  =
  \frac{2 r_s^2 \sqrt{L_\obs^2 + r_\obs^2}}{r_\obs^2}
  \int_0^\theta
  \sin^4 \! \tfrac{\theta}{2}
  \, \dd \theta
\nonumber
\\
  &=
  \frac{r_s^2 \sqrt{L_\obs^2 + r_\obs^2}}{r_\obs^2}
  \bigl(
  \tfrac{3}{4}
  \theta
  -
  \sin\theta
  +
  \tfrac{1}{8}
  \sin 2\theta
  \bigr)
  \ .
\end{align}
The $\theta$-dependent factor in parentheses on the right-hand side of
equation~(\ref{lambdashortestcausal}) is
$\approx \frac{1}{40} \theta^5$ at small separations $\theta$,
and equal to $\tfrac{3}{4} \pi$ at $\theta \rightarrow \pi$.

It should be emphasized that the affine distance $\lambda_\obs$
is a perceived observer-dependent distance.
To fix the normalization would require making a gauge choice of observer.
One canonical choice might be an observer radially infalling at the
horizon, where $J \rightarrow \infty$ light rays begin their journey,
$L_\obs = 0$ and $r_\obs = r_s$,
in which case the observer-dependent factor in
equation~(\ref{lambdashortestcausal}) reduces to $r_s$.
Another possible choice might an observer radially infalling near
the singular surface,
in which case the observer-dependent factor is $r_s^2 / r_\obs$,
which diverges as $r_\obs \rightarrow 0$.

The conclusion is that, although a causal distance between
points on the singular surface can be defined,
that distance cannot be construed as a Lorentz-invariant
measure of separation between points on the singular surface.

\section{Geodesics in the Kerr-Newman geometry}
\label{kngeodesics-sec}

\subsection{Kerr-Newman line-element}

As first pointed out by Carter \cite{Carter:1968c},
geodesics in the Kerr-Newman \cite{Kerr:1963,Newman:1965}
geometry are Hamilton-Jacobi separable.
Although it is most common to express the Kerr-Newman line-element in terms of
Boyer-Lindquist \cite{Boyer:1967}
ellipsoidal coordinates $t$, $r$, $\theta$, $\phi$,
separation of variables is most natural
with respect to a different set of radial and angular coordinates $x$ and $y$
in place of the Boyer-Lindquist radius $r$ and polar angle $\theta$,
equations~(\ref{xrytheta}).
In terms of coordinates
$t$, $x$, $y$, $\phi$
adapted to facilitate separation
\cite{Hamilton:2010b},
the line-element for Kerr-Newman and related separable geometries is
\begin{align}
\label{seplineelement}
  \dd s^2
  =
  \rho^2
  &
  \biggl[
  - \,
  \Delta_x
  \left( {\dd t - \omega_y \, \dd \phi \over 1 - \omega_x \omega_y} \right)^2
  +
  {\dd x^2 \over \Delta_x}
\nonumber
\\
  &\ 
  +
  {\dd y^2 \over \Delta_y}
  +
  \Delta_y
  \left( {\dd \phi - \omega_x \, \dd t \over 1 - \omega_x \omega_y} \right)^2
  \biggr]
  \ .
\end{align}
Separability requires that
$\omega_x$ and $\Delta_x$
are functions only of the radial coordinate $x$,
while
$\omega_y$ and $\Delta_y$
are functions only of the polar coordinate $y$.
The conformal factor $\rho^2$ must be a separated sum of functions of
$x$ and $y$.
The Einstein equations that follow from the line-element~(\ref{seplineelement})
prove to be separable \cite{Hamilton:2010b}.
Solving the Einstein equations
with zero source of energy-momentum,
or with energy-momentum sourced by a central charge,
yields the Kerr-Newman geometry
for a rotating black hole of mass $M$, charge $Q$, and spin $a$.
For the Kerr-Newman geometry, the radial and angular coordinates $x$ and $y$
are related to the Boyer-Lindquist radius $r$ and polar angle $\theta$ by
\begin{equation}
\label{xrytheta}
  x = \frac{1}{a} \arctan\frac{a}{r}
  \ , \quad
  y = - \cos\theta
  \ ,
\end{equation}
whose differentials are related by
\begin{equation}
  \dd x = - \frac{\dd r}{R^2}
  \ , \quad
  \dd y = \sin\theta \, \dd \theta
  \ ,
\end{equation}
with $R$ defined by
\begin{equation}
  R \equiv \sqrt{r^2 + a^2}
  \ .
\end{equation}
For the Kerr-Newman geometry,
the conformal factor $\rho^2$ in the line-element~(\ref{seplineelement})
is the separated sum
\begin{equation}
  \rho^2 = r^2 + a^2 \cos^2\!\theta
  \ ,
\end{equation}
$\omega_x$ is the angular velocity of the Boyer-Lindquist tetrad frame
through the coordinates,
$\omega_y$ is the angular momentum of principal null geodesics,
\begin{equation}
  \omega_x
  =
  {a \over R^2}
  \ , \quad
  \omega_y
  =
  a \sin^2\!\theta
  \ ,
\end{equation}
$\Delta_x$ is the horizon function, and $\Delta_y$ the polar function,
\begin{equation}
\label{Deltaxy}
  \Delta_x
  =
  \frac{1}{R^2}
  \left(
  1 - {2 M r + Q^2 \over R^2}
  \right)
  \ , \quad
  \Delta_y
  =
  \sin^2\!\theta
  \ .
\end{equation}
The vanishing of the horizon function $\Delta_x$ defines the
radial location of
outer and inner horizons $r_+$ and $r_-$, at
\begin{equation}
\label{rpm}
  r_\pm
  =
  M \pm \sqrt{M^2 - Q^2 - a^2}
  \ .
\end{equation}
The vanishing of the polar function $\Delta_y$ defines the location
of north and south poles, at $\theta = 0$ and $\pi$.
Ergospheres, boundaries of regions within which the black hole's rotation
drags around frames faster than light, occur where $g_{tt} = 0$, at
\begin{equation}
\label{epm}
  e_\pm
  =
  M \pm \sqrt{M^2 - Q^2 - a^2 \cos^2\!\theta}
  \ .
\end{equation}

\subsection{Geodesics from Hamilton-Jacobi separation}

The Hamilton-Jacobi equation governing geodesics
of a particle of rest mass $m$
(and zero charge, if the black hole is charged;
see \cite{Hamilton:2010b} for generalization to a charged particle)
is
\begin{equation}
\label{HJeq}
  g^{\mu\nu}
  p_\mu p_\nu
  =
  -
  m^2
  \ ,
\end{equation}
where the covariant momenta
$p_\mu$ are, in the Hamilton-Jacobi approach,
defined to be the derivatives of the particle action $S$
with respect to the coordinates,
\begin{equation}
  p_\mu
  \equiv
  {\partial S \over \partial x^\mu}
  \ .
\end{equation}
In geometries such as Kerr that are time-translation symmetric
and azimuthally symmetric,
the covariant time and azimuthal momenta
$p_t$ and $p_\phi$
define the conserved energy $E$
and azimuthal angular momentum $L$ of the particle,
\begin{equation}
  p_t
  =
  - E
  \ , \quad
  p_\phi
  =
  L
  \ .
\end{equation}
The Hamilton-Jacobi equation~(\ref{HJeq}) is separable if the
the action $S$ is a separated sum of functions of the coordinates
\cite{Carter:1968c},
\begin{equation}
  S
  =
  -\,
  E t
  +
  L \phi
  +
  S_x
  +
  S_y
  \ ,
\end{equation}
in which $S_x$ and $S_y$ are respectively functions of $x$ and $y$ alone.
The line-element~(\ref{seplineelement})
defines not only a metric, but also a tetrad,
a foursome of locally inertial axes at each point,
encoded in a vierbein matrix $e^m{}_\mu$,
\begin{equation}
  \dd s^2
  =
  \eta_{mn}
  e^m{}_\mu
  e^n{}_\nu
  \,
  \dd x^\mu \dd x^\nu
  \ ,
\end{equation}
where $\eta_{mn} \equiv \diag \{ -1 , 1 , 1 , 1 \}$
is the Minkowski metric of special relativity.
Latin and Greek indices denote respectively tetrad and coordinate indices.
Recast into the tetrad frame,
the Hamilton-Jacobi equation~(\ref{HJeq}) is
\begin{equation}
\label{HJtetrad}
  \eta^{mn}
  p_m p_n
  =
  -
  m^2
  \ ,
\end{equation}
in which the covariant tetrad-frame momenta $p_m = e_m{}^\mu p_\mu$
of the particle are
\begin{equation}
\label{momentumtetrad}
  p_m
  =
  {1 \over \rho}
  \left\{
  {P_t \over \sqrt{| \Delta_x |}}
  , \,
  {P_x \over \sqrt{| \Delta_x |}}
  , \,
  {P_y \over \sqrt{\Delta_y}}
  , \,
  {P_\phi \over \sqrt{\Delta_y}}
  \right\}
  \ ,
\end{equation}
with Hamilton-Jacobi parameters $P_m$ defined in terms of
the coordinate-frame momenta $p_\mu$ by
\begin{subequations}\label{Ptphixy}
\begin{alignat}{3}
\label{Ptphi}
  P_t
  &=
  %p_t + p_\phi \omega_x
  %=
  -\, E + L \omega_x
  \ , \quad
  &P_\phi
  &=
  %p_\phi + p_t \omega_y
  %=
  L - E \omega_y
  \ ,
\\
\label{Pxy}
  P_x
  &=
  - p_x \Delta_x
  \ , \quad
  &P_y
  &=
  p_y \Delta_x
  \ .
\end{alignat}
\end{subequations}
In terms of the Hamilton-Jacobi parameters $P_m$,
the Hamilton-Jacobi equation~(\ref{HJtetrad}) is
\begin{equation}
\label{HamiltonJacobisep}
  {- \, P_t^2 + P_x^2
  \over \Delta_x}
  +
  {P_y^2 + P_\phi^2
  \over \Delta_y}
  =
  -
  m^2 \rho^2
  \ .
\end{equation}
The Hamilton-Jacobi equation~(\ref{HamiltonJacobisep}) separates as
\begin{equation}
\label{HamiltonJacobisepK}
  -
  \left(
  {- \, P_t^2 + P_x^2
  \over \Delta_x}
  +
  m^2 r^2
  \right)
  =
  {P_y^2 + P_\phi^2
  \over \Delta_y}
  +
  m^2
  a^2 \cos^2\!\theta
  =
  \KCarter
  \ ,
\end{equation}
where $\KCarter$ is a separation constant, the Carter constant.
Physically, the Carter constant $\KCarter$ is a constant of motion
associated with conservation of total angular momentum squared.
The separated Hamilton-Jacobi equations~(\ref{HamiltonJacobisepK}) imply that
\begin{subequations}
\label{PUxy}
\begin{align}
\label{PUr}
  P_x
  &=
  \pm
  \sqrt{
  P_t^2
  - \left(
  \KCarter
  +
  m^2 r^2
  \right)
  \Delta_x
  }
  \ ,
\\
\label{PUtheta}
  P_y
  &=
  \pm
  \sqrt{
  - P_\phi^2
  +
  \left(
  \KCarter
  -
  m^2 a^2 \cos^2\!\theta
  \right)
  \Delta_y
  }
  \ .
\end{align}
\end{subequations}
The expressions~(\ref{PUxy}) for the Hamilton-Jacobi parameters $P_x$ and $P_y$
can be interpreted as effective radial and polar potentials
governing the motion of the particle in the radial $x$ and polar $y$ directions.
The coordinates along a particle geodesic follow from integrating
$m \, \dd x^{\mu} / \dd \tau = p^\mu = e_m{}^{\mu} p^k$,
which in the present case involves integrating
$\dd y / \dd x = p^y / p^x = - P_y / P_x$,
equivalent to the implicit equation
\begin{equation}
\label{dxdyP}
  -
  {\dd x \over
  P_x}
%  \sqrt{
%  P_t^2
%  - \left(
%  \KCarter
%  +
%  m^2 \rhox^2
%  \right)
%  \Delta_x
%  }}
  =
  {\dd y \over
  P_y}
%  \sqrt{
%  - P_\phi^2
%  +
%  \left(
%  \KCarter
%  -
%  m^2 \rhoy^2
%  \right)
%  \Delta_y
%  }}
  \ .
\end{equation}
The time and azimuthal coordinates $t$ and $\phi$ along the trajectory
follow from integrating
\begin{subequations}
\begin{align}
\label{dtdphiP}
  \dd t
  &=
  {P_t \, \dd x \over P_x \Delta_x}
  +
  {\omega_y P_\phi \, \dd y \over P_y \Delta_y}
  \ ,
\\
  \dd \phi
  &=
  {\omega_x P_t \, \dd x \over P_x \Delta_x}
  +
  {P_\phi \, \dd y \over P_y \Delta_y}
  \ .
\end{align}
\end{subequations}

%Note that
%\begin{equation}
%\label{dphidoranP}
%  \dd \phi - \omega_x \, \dd t
%  =
%  {P_\phi ( 1 - \omega_x \omega_y ) \, \dd y \over P_y \Delta_y}
%  \ .
%\end{equation}

\subsection{Conditions on Hamilton-Jacobi parameters}

The Hamilton-Jacobi equation~(\ref{HamiltonJacobisep}) can be rearranged as
\begin{equation}
\label{HamiltonJacobihor}
  P_t^2 - P_x^2
  =
  \left(
  {P_y^2 + P_\phi^2
  \over \Delta_y}
  +
  m^2 \rho^2
  \right)
  \Delta_x
  \ ,
\end{equation}
which shows that $P_t^2 - P_x^2$ must change sign across a horizon,
where the horizon function $\Delta_x$ passes through zero.
Equation~(\ref{HamiltonJacobihor})
shows that the Hamilton-Jacobi parameters $P_t$ and $P_x$ must satisfy
\begin{equation}
\label{Ptxconditions}
  \begin{array}{cl}
  | P_t | > | P_x |
  &
  \mbox{if $\Delta_x > 0$}
  \ ,
  \\
  | P_t | = | P_x |
  &
  \mbox{if $\Delta_x = 0$}
  \ ,
  \\
  | P_t | < | P_x |
  &
  \mbox{if $\Delta_x < 0$}
  \ .
  \end{array}
\end{equation}
The Hamilton-Jacobi parameters $P_m$ must be continuous,
including across horizons.
The conditions~(\ref{Ptxconditions}) imply that
$P_t$ must have the same sign everywhere throughout
any connected region where the horizon function $\Delta_x$ is positive,
while $P_x$ must have the same sign everywhere throughout
any connected region where $\Delta_x$ is negative.
In summary, the signs of the Hamilton-Jacobi parameters $P_t$ and $P_x$
must be as follows.
\begin{enumerate}
\item
Outside the outer horizon, in the Universe part of the Kerr geometry
in Fig.~\ref{penrosemassinflation},
the time parameter $P_t$ must be negative,
reflecting the fact that
the time coordinate $t$ must be timelike and increasing
with the proper time of any particle.
The radial parameter $P_x$ can be either positive (outfalling)
or negative (infalling).
\item
Between the outer and inner horizons, in the Black Hole part of the geometry
in Fig.~\ref{penrosemassinflation},
the radial parameter $P_x$ must be negative,
reflecting the fact that the radius is timelike and decreasing
with the proper time of any particle.
The time parameter $P_t$ can be either positive (outgoing)
or negative (ingoing).
\item
Inside the inner horizon,
the time parameter $P_t$ is
positive in the Outgoing wormhole,
negative in the Ingoing wormhole
part of the geometry in Fig.~\ref{penrosemassinflation},
The radial parameter $P_x$ can be either positive (outfalling)
or negative (infalling).
\end{enumerate}

%\Ptab

%\begin{acknowledgments}
%This paper was supported in part by hope.
%\end{acknowledgments}

%\bibliographystyle[unsrt]
\bibliography{bh}

\end{document}